\documentclass[prd,superscriptaddress,showpacs,nofootinbib,amsmath,amssymb,aps,10pt]{revtex4}

\usepackage{bm}
\usepackage{amsfonts}
\usepackage{latexsym}
\usepackage[latin1]{inputenc}
\usepackage{graphicx}
\usepackage{amsmath}
\usepackage{palatino}
\usepackage{mathpazo}
\usepackage{booktabs}
\usepackage{dcolumn}

\def\jnl@style{\it}
\def\aaref@jnl#1{{\jnl@style#1}}

\def\aaref@jnl#1{{\jnl@style#1}}

\def\aj{\aaref@jnl{AJ}}                   
\def\apj{\aaref@jnl{ApJ}}                 
\def\apjl{\aaref@jnl{ApJ}}                
\def\apjs{\aaref@jnl{ApJS}}               
\def\apss{\aaref@jnl{Ap\&SS}}             
\def\aap{\aaref@jnl{A\&A}}                
\def\aapr{\aaref@jnl{A\&A~Rev.}}          
\def\aaps{\aaref@jnl{A\&AS}}              
\def\mnras{\aaref@jnl{Mon.~Not.~Roy.~Astron.~Soc.}}             
\def\prd{\aaref@jnl{Phys.~Rev.~D}}        
\def\prc{\aaref@jnl{Phys.~Rev.~C}}  
\def\prl{\aaref@jnl{Phys.~Rev.~Lett.}}    
\def\qjras{\aaref@jnl{QJRAS}}             
\def\skytel{\aaref@jnl{S\&T}}             
\def\ssr{\aaref@jnl{Space~Sci.~Rev.}}     
\def\zap{\aaref@jnl{ZAp}}                 
\def\nat{\aaref@jnl{Nature}}              
\def\aplett{\aaref@jnl{Astrophys.~Lett.}} 
\def\apspr{\aaref@jnl{Astrophys.~Space~Phys.~Res.}} 
\def\physrep{\aaref@jnl{Phys.~Rep.}}      
\def\physscr{\aaref@jnl{Phys.~Scr}}       
\def\commat{\aaref@jnl{Comm.~Math.~Phys.}}              
\def\science{\aaref@jnl{Science}}               
\def\cqg{\aaref@jnl{Classical Quant.~Grav.}}            
\def\jpcs{\aaref@jnl{JPCS}}                                     
\def\ijmpd{\aaref@jnl{Int.~J.~Mod.~Phys.~D}}                    
\def\grg{\aaref@jnl{Gen.~Relat.~Gravit.}}               
\def\rpp{\aaref@jnl{Rep.~Prog.~Phys.}}          
\def\npa{\aaref@jnl{Nucl.~Phys.~A}}        
\def\lrr{\aaref@jnl{Living Rev.~Rel.}}                   
\def\jcap{\aaref@jnl{J.~Cosmology Astropart.~Phys.}}    
\def\rmp{\aaref@jnl{Rev.~Mod.~Phys.}}   

\allowdisplaybreaks[1]
\renewcommand{\arraystretch}{1.1}
\usepackage{color}

\begin{document}

\title{Orbital and epicyclic frequencies around rapidly rotating compact stars in scalar-tensor theories of gravity}

\author{Daniela D. Doneva}
\email{daniela.doneva@uni-tuebingen.de}
\affiliation{Theoretical Astrophysics, Eberhard Karls University of T\"ubingen, T\"ubingen 72076, Germany}
\affiliation{INRNE - Bulgarian Academy of Sciences, 1784  Sofia, Bulgaria}

\author{Stoytcho S. Yazadjiev}
\affiliation{Department of Theoretical Physics, Faculty of Physics, Sofia University, Sofia 1164, Bulgaria}
\affiliation{Theoretical Astrophysics, Eberhard Karls University of T\"ubingen, T\"ubingen 72076, Germany}

\author{Nikolaos Stergioulas}
\affiliation{Department of Physics, Aristotle University of Thessaloniki, Thessaloniki 54124, Greece}

\author{Kostas D. Kokkotas}
\affiliation{Theoretical Astrophysics, Eberhard Karls University of T\"ubingen, T\"ubingen 72076, Germany}
\affiliation{Department of Physics, Aristotle University of Thessaloniki, Thessaloniki 54124, Greece}

\author{Tilemachos M. Athanasiadis}
\affiliation{Department of Physics, Aristotle University of Thessaloniki, Thessaloniki 54124, Greece}


\begin{abstract}
 We study the orbital and epicyclic frequencies of particles orbiting around rapidly rotating neutron stars and strange stars in a particular scalar-tensor theory of gravity.  We find very large deviations of these frequencies, when compared to their corresponding values in general relativity, for the maximum-mass rotating models. \ In contrast, for models rotating with spin frequency of 700Hz (approximately the largest known rotation rate of neutron stars), the deviations are generally small. Nevertheless, for a very stiff equation of state and a high mass the deviation of one of the epicyclic frequencies from its GR value is appreciable even at a spin frequency of 700Hz. In principle, such a deviation could become important in models of quasi-periodic oscillations in low-mass x-ray binaries and could serve as a test of strong gravity (if other parameters are well constraint). Even though the present paper is concentrated mainly on orbital and epicyclic frequencies, we present here for the first time rapidly rotating, scalarized equilibrium compact stars with realistic hadronic equations of state and strange matter equation of state.  We also provide analytical expressions for the exterior spacetime of scalarized neutron stars and their epicyclic frequencies in the nonrotating limit.
\end{abstract}


\maketitle

\section{Introduction}
Accretion in Low-Mass X-ray Binaries (LMXBs) takes place in a region of the spacetime around compact objects where strong gravity effects play an important role. In many cases, the presence of quasiperiodic oscillations (QPOs) with frequencies in the Hz to kHz range  has been detected.  Various approaches have been used for explaining the QPOs: relativistic precession models, relativistic resonance models, beat frequency models, preferred radii models and others (for a review see e.g. \cite{Klis2006}). Most of these models are related in one way or another to the three characteristic frequencies of particles orbiting around compact objects, namely the orbital frequency and the radial and vertical epicyclic frequencies. On the other hand, a different approach has been taken in \cite{Rezzolla2003,Rezzolla2003a,Montero2004} where the global oscillations of thick accretion discs (accretion torii) were considered, but some of the oscillations of tori are modified (due to finite pressure effects)\ epicyclic frequencies.

The fact that QPOs originate in a strong gravity regime, naturally leads us to the idea of using them as a test of alternative theories of gravity, since almost all of the current observations constraining the alternative theories of gravity are in the weak field regime (see, e.g. \cite{Will2006}). One of the simplest and most natural generalizations of General Relativity (GR) are the scalar-tensor theories of gravity (STTs). Their essence is in one or several scalar fields which are mediators of the gravitational interaction, in addition to the spacetime metric. A very important property of STT is that in the physical Jordan frame there is no direct interaction between the matter and the scalar field, and consequently the weak equivalence principle is satisfied.

Different proposals of STTs have been examined in the literature and an interesting subclass are cases which are indistinguishable from GR in the weak field regime, but which can have appreciable deviations in strong fields. In such cases interesting phenomena, such as non-uniqueness and bifurcations of the solutions, could exist. An example is the so-called spontaneous scalarization of neutron stars \cite{Damour1993}. In a density regime near the maximum mass for neutron stars, new solutions that have a nontrivial scalar field appear, in addition to the trivial (GR) solution. It turns out that the scalarized solution is energetically more favorable \cite{Damour1993} than the GR solution and its stability was studied in \cite{Harada1998,Harada1997}. Different astrophysical implications of scalarized neutron stars were examined in \cite{Sotani04,Sotani2005,DeDeo2003,DeDeo2004,Barausse2013,Palenzuela2014,Shibata2014,Novak1998,Novak1998a,Harada1997a,Sotani2014}. Similar nonlinear phenomena in STTs are present also in the black hole case \cite{Stefanov2008,Doneva2010,Pani2011a,Cardoso2013a}.

Most studies of neutron stars in STT were for either nonrotating models or in the   slow rotation approximation (of linear order in the angular velocity) \cite{Damour1996,Sotani2012}. The only exception is our recent work \cite{Doneva2013} where we obtained equilibrium models of scalarized, rapidly rotating neutron stars, up to the mass-shedding limit. In    \cite{Doneva2013} we  found that rapidly rotating scalarized neutron stars exist for a significantly wider range of central densities, compared to the nonrotating case. In addition, the mass and radius of rapidly rotating scalarized models differ significantly more from their GR counterparts, than in the nonrotating case. Thus, even for moderate values of the scalar-field coupling parameter, for which scalarized nonrotating solutions either do not exist or differ only marginally from the GR solutions, rapidly rotating scalarized neutron stars exist and can deviate considerably from neutron stars in  pure Einstein's theory.

The effect of the scalar field on the epicyclic frequency (for a particular version of STT) was examined in \cite{DeDeo2004} for the case of nonrotating neutron stars. Accreting neutron stars in LMXBs can be rapidly rotating, with spin frequencies reaching approximately $700$Hz \cite{Hessels2006} and thus we present here the first study of epicyclic frequencies around rapidly rotating scalarized neutron stars. We find that although the effect of a nontrivial scalar field on the epicyclic frequencies is quite small for nonrotating models (for values of the coupling parameter that are in agreement with the observational constraints \cite{Freire2012,Antoniadis13}), it is more appreciable for rapidly rotating stars.

The paper is organized as follows:\ In Section II we give the theoretical background and the equations for the orbital and epicyclic frequencies. The results are presented in Section III in the case of neutron and strange stars. In the Appendix we  derive analytical formulae for the orbital and epicyclic frequencies in the nonrotating case.

\section{MAIN equations} \label{Sec:BasicEquation}
The action in scalar-tensor theories of gravity, in the physical \textit{Jordan} frame, is given by \cite{Fujii2003,Damour1992}:
\begin{eqnarray} \label{JFA}
S = {1\over 16\pi G_{*}} \int d^4x \sqrt{-{\tilde
g}}\left[{F(\Phi)\tilde R} - Z(\Phi){\tilde
g}^{\mu\nu}\partial_{\mu}\Phi
\partial_{\nu}\Phi   -2 U(\Phi) \right] +
S_{m}\left[\Psi_{m};{\tilde g}_{\mu\nu}\right] ,
\end{eqnarray}
where $G_{*}$ is the bare gravitational constant, $\Phi$ is the scalar field, ${\tilde R}$ is the Ricci scalar curvature with respect to the spacetime metric ${\tilde g}_{\mu\nu}$ , ${\tilde g}$ is the determinant of the metric and $F, Z, U$ are functions of the scalar field.  The matter fields are collectively denoted by $\Psi_{m}$ and their action is $S_{m}$. As one can notice, the scalar field does not appear explicitly in the action of the matter, in order for the weak equivalence principle to be satisfied\footnote{Of course one can consider more general cases with a direct coupling between the matter and the scalar field, but this is out of the scope of the current paper.}. Instead, the scalar field influences the matter only through the spacetime metric. This means that the equations of motion of test particles in scalar-tensor theories are the same as the equations of motion in general relativity.

Having this in mind and using the procedure described in \cite{Ryan1995,Pappas2012b}, one can easily derive the orbital and epicyclic frequencies of a particle orbiting around a compact object by examining the geodesics and their perturbations. For rapidly rotating neutron stars we consider a stationary, axisymmetric spacetime (see \cite{Friedman2013} for the general form of the metric). It is easy to show that in the usual quasi-isotropic coordinates, the orbital frequency for a circular equatorial orbit with coordinate radius $r_c$ is given by:
\begin{eqnarray}\label{Eq:OmegaP}
\Omega_{p} = \left.\frac{-\partial_r {\tilde g}_{t\phi} \pm \sqrt{(\partial_r {\tilde g}_{t\phi})^2 - \partial_r {\tilde g}_{tt}\partial_r {\tilde g}_{\phi\phi}}}{\partial_r {\tilde g}_{\phi\phi}}\right|_{r=r_c, \theta=\frac{\pi}{2}},
\end{eqnarray}
where ${\tilde g}_{tt}$, ${\tilde g}_{\phi\phi}$ and ${\tilde g}_{t\phi}$ are the corresponding components of the metric.

If a particle on a stable circular orbit is perturbed, then it oscillates with some characteristic epicyclic frequencies $\omega_r$ and $\omega_\theta$ in the radial or vertical direction, respectively. These frequencies can be obtained by perturbing the equations of motion and after some calculations we arrive at \cite{Ryan1995,Pappas2012b}:
\begin{eqnarray}
\omega_r^2 &=& \frac{1}{2{\tilde g}_{rr}}\left[({\tilde g}_{tt}+{\tilde g}_{t\phi}\Omega_p)^2\;\partial_r^2 \left(\frac{{\tilde g}_{\phi\phi}}{Y}\right) -2 ({\tilde g}_{tt}+{\tilde g}_{t\phi}\Omega_p)({\tilde g}_{t\phi}+{\tilde g}_{\phi\phi}\Omega_p) \;\partial_r^2 \left(\frac{{\tilde g}_{t\phi}}{Y}\right) \right. + \notag \\ \notag \\
&&\left.+ ({\tilde g}_{t\phi}+{\tilde g}_{\phi\phi}\Omega_p)^2\;\partial_r^2 \left(\frac{{\tilde g}_{tt}}{Y}\right)\right]_{r=r_c, \theta=\frac{\pi}{2}},
\label{eq:epicrad}\\ \notag \\
\omega_\theta^2 &=& \frac{1}{2{\tilde g}_{\theta\theta}}\left[({\tilde g}_{tt}+{\tilde g}_{t\phi}\Omega_p)^2\;\partial_\theta^2 \left(\frac{{\tilde g}_{\phi\phi}}{Y}\right) -2 ({\tilde g}_{tt}+{\tilde g}_{t\phi}\Omega_p)({\tilde g}_{t\phi}+{\tilde g}_{\phi\phi}\Omega_p) \;\partial_\theta^2 \left(\frac{{\tilde g}_{t\phi}}{Y}\right) \right. + \notag \\ \notag \\
&&\left.+ ({\tilde g}_{t\phi}+{\tilde g}_{\phi\phi}\Omega_p)^2\;\partial_\theta^2 \left(\frac{{\tilde g}_{tt}}{Y}\right)\right]_{r=r_c, \theta=\frac{\pi}{2}},
\label{eq:epicvert}
\end{eqnarray}
where
\begin{eqnarray}
Y = {{\tilde g}_{tt}{\tilde g}_{\phi\phi}-\tilde g}_{t\phi}^2 .
\end{eqnarray}
Eqns. (\ref{eq:epicrad}) and (\ref{eq:epicvert}) for the epicyclic frequencies of particles orbiting a compact object  are the same in STT as in GR.

For a nonrotating star the vertical epicyclic frequency is equal to the orbital frequency, i.e. $\omega_\theta = \Omega_p$ for $\Omega=0$, where $\Omega$ is the angular velocity of the star, while  $\omega_r^2$ can vanish at the innermost stable circular orbit (ISCO) near the neutron star surface, signaling radial instability of the circular orbits at smaller radii (an effect absent in Newtonian gravity for spherical, ideal fluid sources).  In simple accretion disk models, the ISCO defines the inner edge of the disc. For some neutron star models,  $\omega_r^2$ does not become negative outside the stellar surface so that all circular orbits are stable (the accretion disc can extend down to the surface).

While the Jordan frame is the physical frame in which observable quantities are measured,
the field equations take a somewhat simpler form in the so-called \textit{Einstein} frame, where the metric $g_{\mu\nu}$ is related to the physical metric  $\tilde g_{\mu\nu}$
by the conformal transformation
\begin{equation}\label {CONF1}
g_{\mu\nu} = F(\Phi){\tilde g}_{\mu\nu} .
\end{equation}
If one introduces a new scalar field $\varphi$ (the dilaton field) satisfying
\begin{equation}\label {CONF2}
\left(d\varphi \over d\Phi \right)^2 = {3\over
4}\left({d\ln(F(\Phi))\over d\Phi } \right)^2 + {Z(\Phi)\over 2
F(\Phi)},
\end{equation}
then the inverse transformation can be written as
\begin{equation}\label {CONF4}
{\tilde g}_{\mu\nu} = {\cal A}^2(\varphi) g_{\mu\nu}
\end{equation}
and the Einstein frame action has the following simpler form:
\begin{eqnarray}
S= {1\over 16\pi G_{*}}\int d^4x \sqrt{-g} \left(R -
2g^{\mu\nu}\partial_{\mu}\varphi \partial_{\nu}\varphi -
4V(\varphi)\right)+ S_{m}[\Psi_{m}; {\cal A}^{2}(\varphi)g_{\mu\nu}],
\end{eqnarray}
where
\begin{eqnarray}\label {CONF3}
{\cal A}(\varphi) &=& F^{-1/2}(\Phi), \\
V(\varphi) &=& U(\Phi)F^{-2}(\Phi)/2.
\end{eqnarray}

The complication that arises is that in this frame the scalar field appears explicitly in the action of the matter via the coupling function ${\cal A}(\varphi)$. Nevertheless, the field equations are simpler in the Einstein frame so that has become the preferred frame for solving the field equations to construct  models of neutron stars or black holes in STT.

In order to obtain the solutions describing scalarized rapidly rotating neutron stars, we used a modification of the {\tt rns} code \cite{Stergioulas95} implemented recently in \cite{Doneva2013}. In \cite{Doneva2013} only the case of a polytropic equation of state (EOS) was considered -- here we use a sample of tabulated, microphysical EOSs.
More details on the field equations governing rapidly rotating neutron stars in STT and on their properties such as mass, radius, angular momentum, etc., can be found in \cite{Doneva2013}.
After obtaining the numerical solution describing a neutron star equilibrium model in the Einstein frame, the physical metric is obtained from Eq. \eqref{CONF4} and used  in Eqs. \eqref{Eq:OmegaP}, (\ref{eq:epicrad}) and (\ref{eq:epicvert}) for evaluating the orbital and epicyclic frequencies.

We consider the case of a vanishing scalar field potential, $V(\varphi)=0$, and choose the following standard form of the  coupling function:
\begin{equation}
{\cal A}(\varphi) = e^{\frac{1}{2} \beta \varphi^2},
\end{equation}
which leads to a scalar-tensor theory that is indistinguishable from GR in the weak field regime, but which can differ significantly when strong fields are considered. More precisely, for negative values of $\beta $, spontaneous scalarization of the neutron stars can be observed for a density range near the maximum mass model \cite{Damour1993,Damour1996,Doneva2013} -- the scalarized solution exists in addition to the (trivial) GR solution with vanishing scalar field, but is energetically favoured over the latter.
Rapid rotation both extends the range of central densities and values of $\beta$ for which scalarization occurs, and significantly enhances the differences in mass and radius with respect to GR.

\section{NUMERICAL\ Results}\label{Sec:Results}

For the EOS, we consider both hadronic and strange matter EOSs. In addition to the properties of orbital and  epicyclic frequencies, we point out that we present the first rapidly rotating, scalarized equilibrium models for tabulated hadronic EOSs and for strange star EOSs\footnote{To our knowledge, scalarization of strange stars has not been studied before even in the nonrotating case.}.

\subsection{Hadronic EOSs}

\begin{figure}[ht!]
\centering
\includegraphics[width=0.48\textwidth]{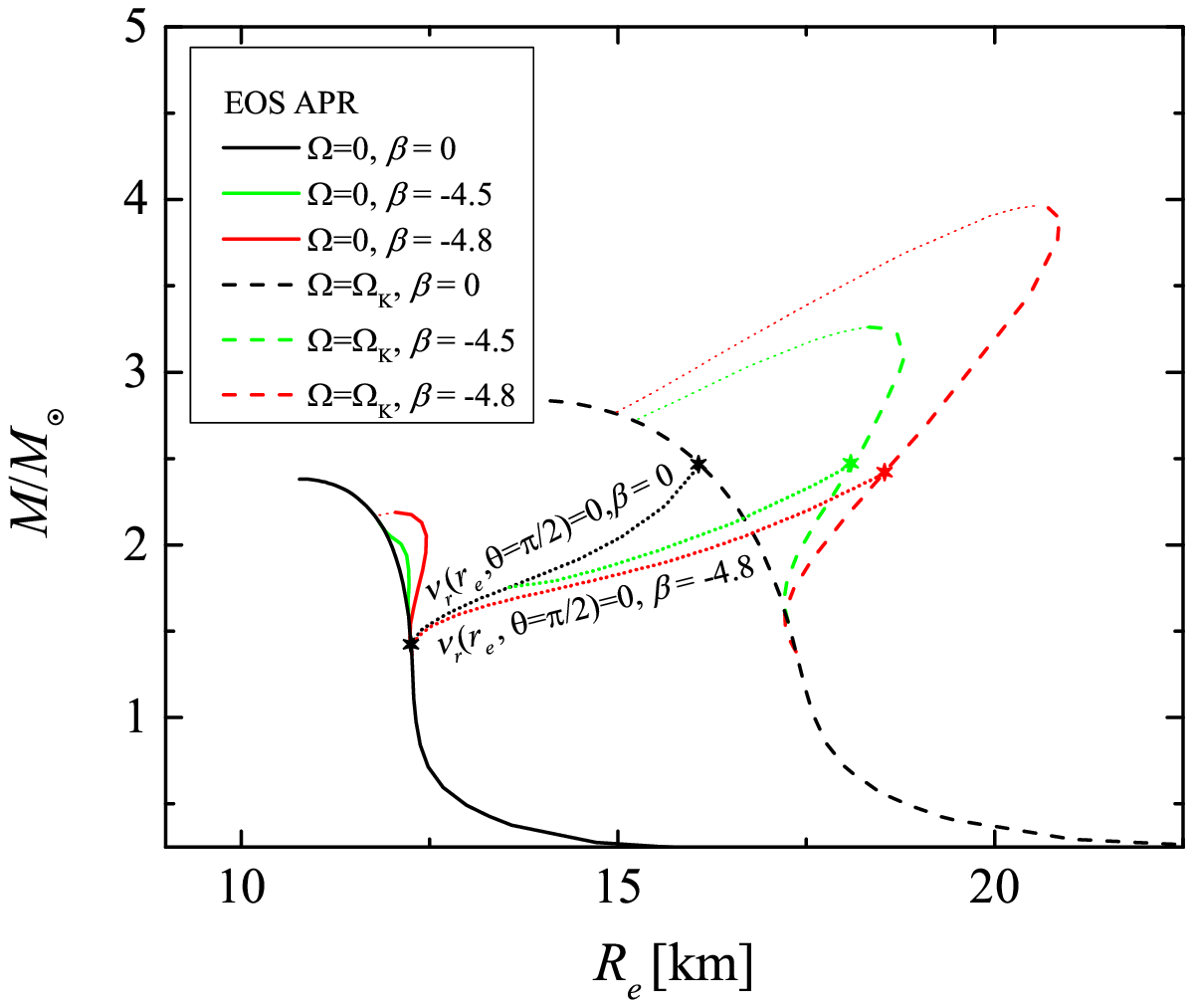}
\includegraphics[width=0.48\textwidth]{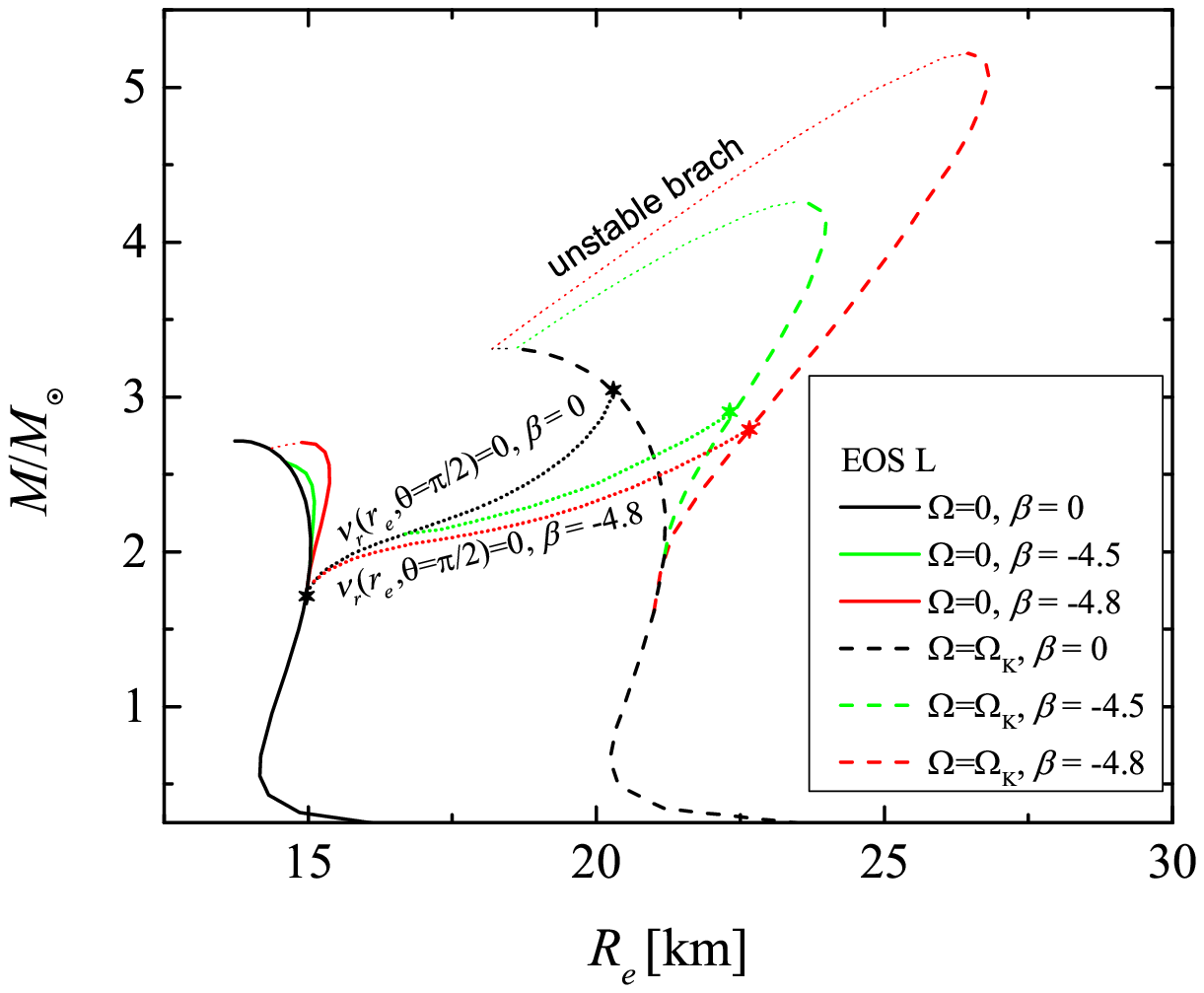}
\caption{The mass as a function of the radius for EOS APR and EOS L, for two different values of $\beta$. The solid lines correspond to nonrotating solutions and the dashed lines to models rotating at the mass-shedding (Kepler) limit (in each case, the branches with models unstable to collapse are shown with thin dotted lines).  Models that possess a gap between the ISCO\ and the surface are above the nearly horizontal dotted lines, while models where all circular orbits in the equatorial plane are stable are below those lines.  }
\label{Fig:M(R)}
\end{figure}

We consider two representative hadronic EOSs that cover a large domain in the presently uncertain mass-radius relation of neutron stars.  EOS APR \cite{AkmalPR} has an average stiffness, while EOS L \cite{Pandharipande1976} is one of the stiffest tabulated  EOSs that have been proposed. Both have a maximum mass larger than the $2 M_\odot$ observational constraint \cite{Demorest10,Manousakis12,Antoniadis13,Lattimer12}. The  mass vs. radius relation for these two EOSs is shown in Fig. \ref{Fig:M(R)}. Two sequences of models are shown -- nonrotating (solid lines) and models rotating up to the mass-shedding (Kepler) limit, where  $\Omega=\Omega_K$ (dashed lines). The unstable (to collapse) branch is shown as a dotted line.\footnote{For the nonrotating models, the instability to collapse sets in at the maximum mass model. We use the same criterion for models at the mass-shedding limit to indicate \textit{approximately} the onset of the quasi-radial instability (the actual marginally stable model will be nearby). }  In each case, we calculate neutron star solutions with $\beta=0$ (the GR case) and solutions with $\beta=-4.5$ and $\beta=-4.8$, which are scalarized in a certain range of central densities (shown in green and red colors) and coincide with the GR solution outside of this range. The current constraint on the coupling parameter from astrophysical observations is $\beta\geq-4.5$ \cite{Freire2012,Antoniadis13}. However, we also include the somewhat stronger case of $\beta=-4.8$ in order to demonstrate the sensitivity of our results on the value of the coupling parameter.

Similarly to the case of polytropes that we studied in  \cite{Doneva2013}, we find that also for the hadronic EOSs rapid rotation both enlarges the range of central densities for which scalarized solutions exist and causes significantly larger deviations from the GR solution, compared to the nonrotating limit. Even though for the current bound of $\beta=-4.5$ the scalarization has only a marginal effect on the structure of nonrotating neutron stars, rapidly rotating models still show a significant effect. In addition to the nonrotating and mass-shedding sequences, Fig. \ref{Fig:M(R)} also shows (with dotted lines) the models for which the corotating ISCO\ touches the surface of the star.  Models above the dotted lines possess a gap between the ISCO and the surface, while for models below the dotted lines there is no region of unstable circular orbits and an accretion disk can reach the surface.  

\begin{figure}[]
\centering
\includegraphics[width=0.48\textwidth]{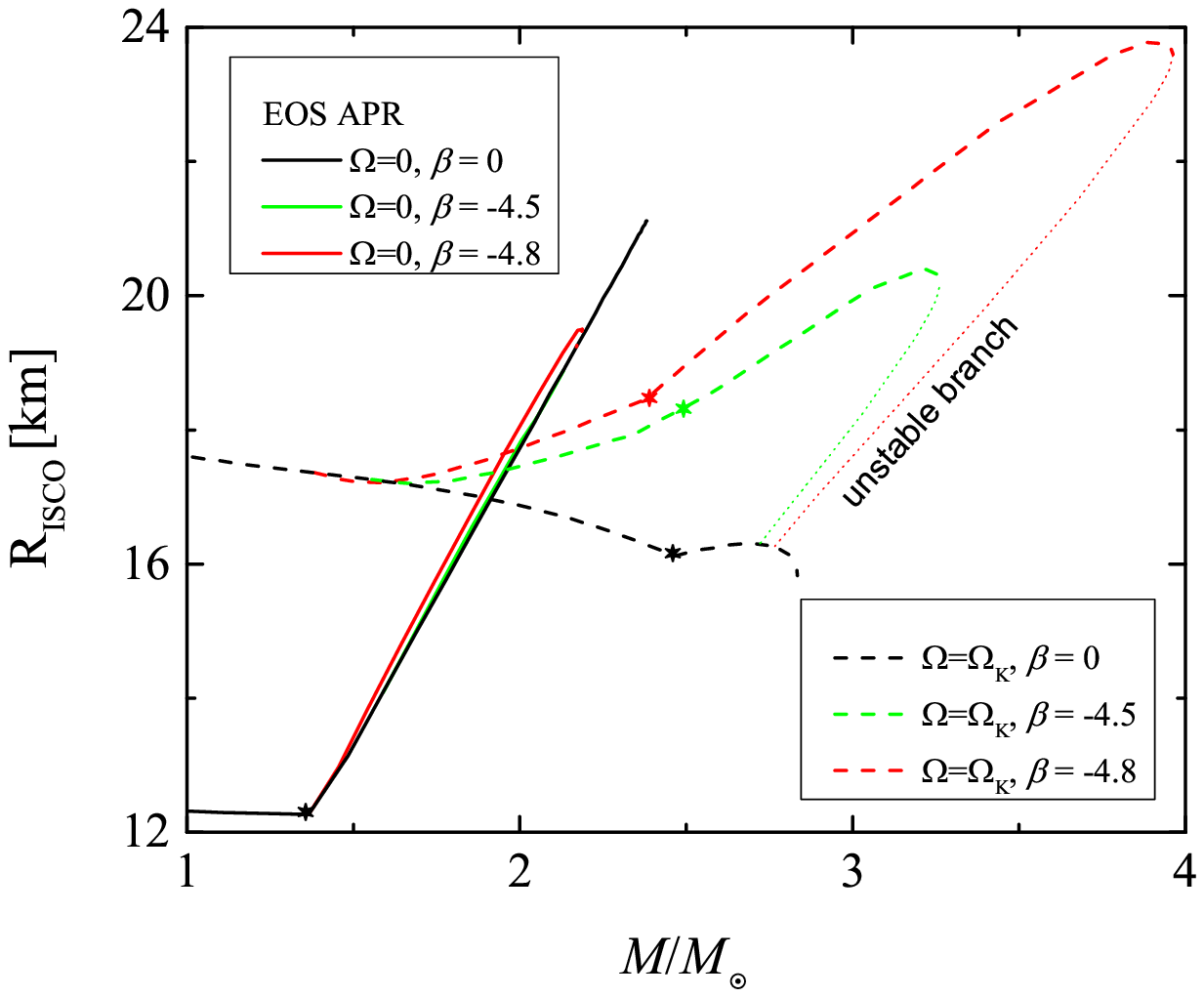}
\includegraphics[width=0.48\textwidth]{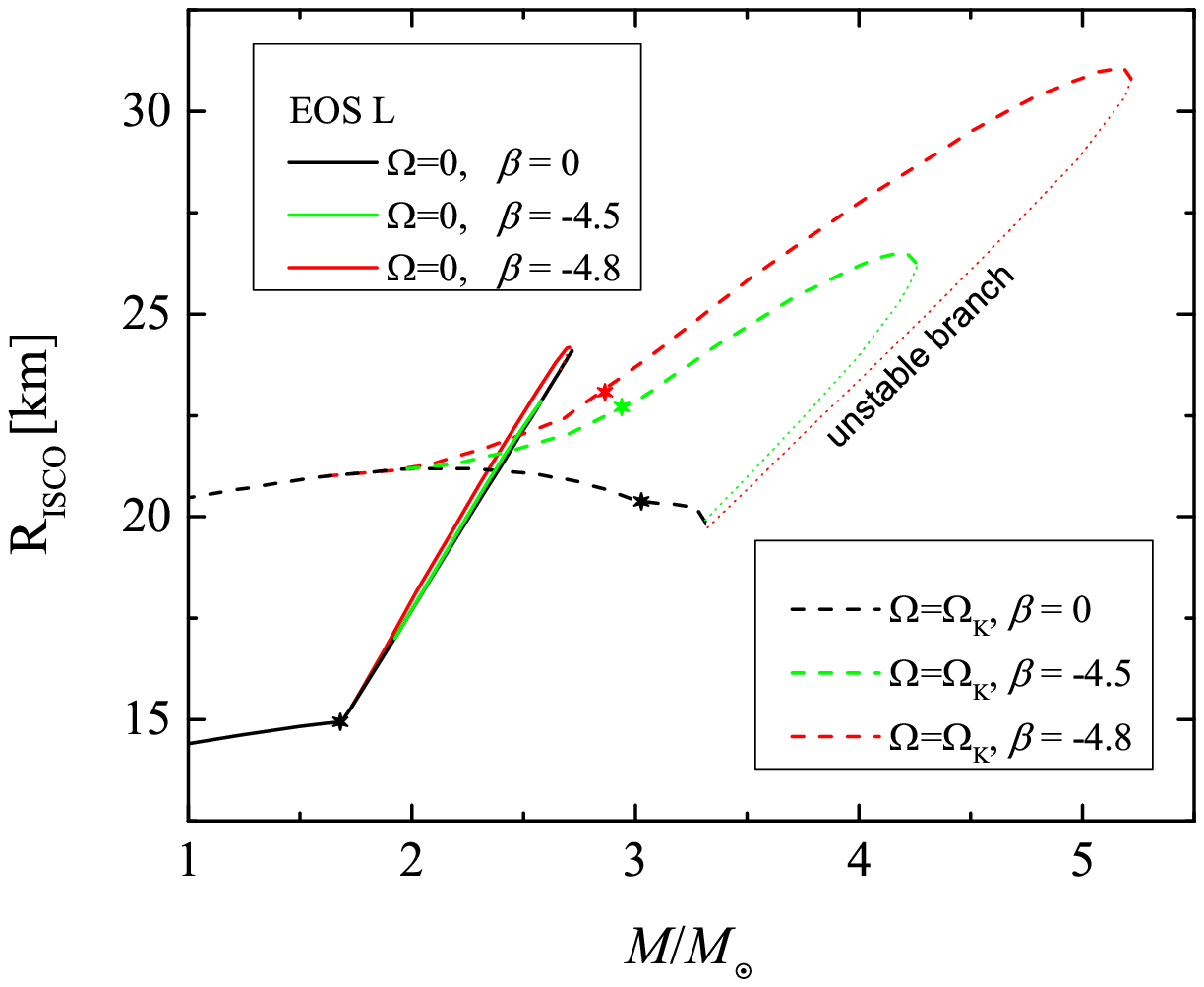}
\caption{The radius of the corotating ISCO for nonrotating models (solid lines) and for models rotating at the mass-shedding limit (dashed lines), as a function of mass along these sequences. An asterisk separates low-mass models, for which there are no unstable circular orbits, from higher-mass models, for which the ISCO is outside the surface of the star.  }
\label{Fig:risco(M)}
\end{figure}

Next, we investigate  the deviations of orbital properties of STT\ solutions from GR solutions in two cases: for nonrotating models and for models rotating at the mass-shedding limit. All of the quantities are plotted as a function of mass along these two sequences. Fig. \ref{Fig:risco(M)} shows the radius of the ISCO. Notice that for low masses, when the ISCO touches the surface, the radius of the surface is shown instead and we indicate the model for which a gap between the ISCO and the surface starts to form with an asterisk. In the nonrotating limit, the ISCO is only marginally affected by scalarization. In contrast, for stars rotating at the mass-shedding limit, the effect is much more pronounced and the radius of the  ISCO for the maximum mass model increases by 25\% for EOS APR and by 33\% for EOS L. Here and below, all quantitative comparisons with the GR solution will be reported for $\beta=-4.5$ (we remind that we show results for $\beta=-4.8$ only as a indication of the sensitivity of the results on the value of $\beta$).

\begin{figure}[]
\centering
\includegraphics[width=0.48\textwidth]{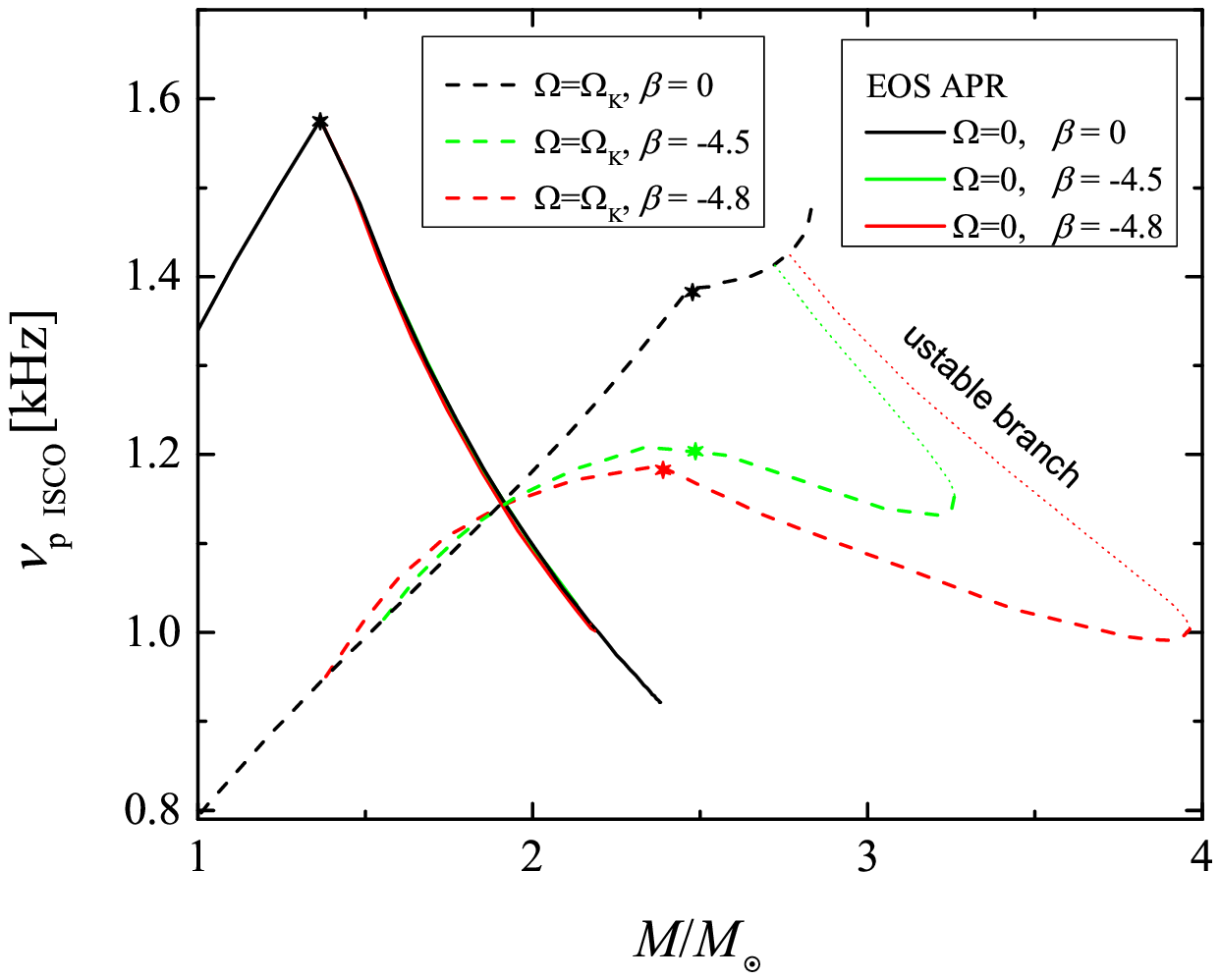}
\includegraphics[width=0.48\textwidth]{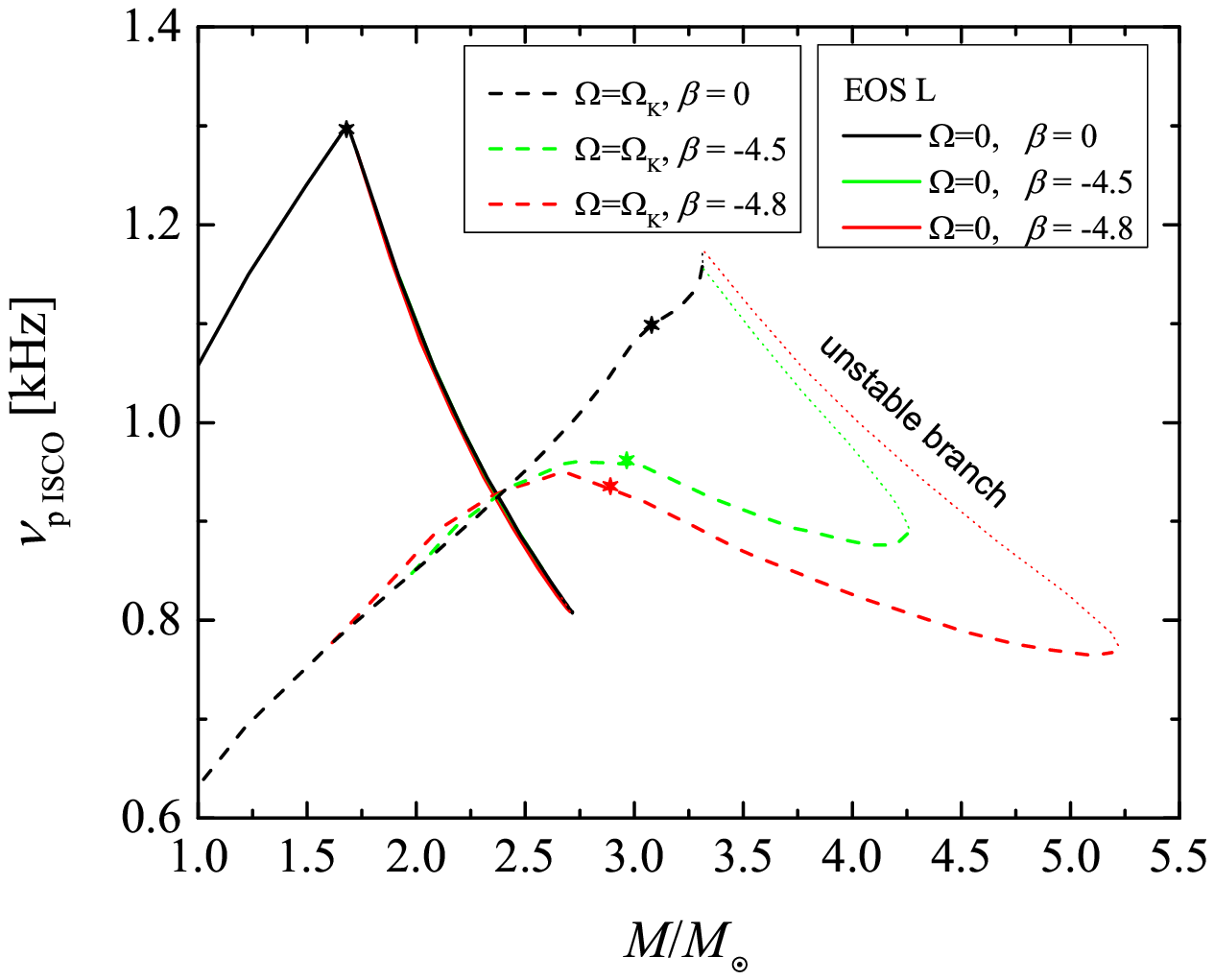}
\caption{The orbital (Keplerian) frequency of a particle at the ISCO (or at the surface for low-mass models) as a function of mass. The notation is the same as in Fig. \ref{Fig:risco(M)}. }
\label{Fig:OmK1(M)}
\end{figure}

\begin{figure}[]
\centering
\includegraphics[width=0.48\textwidth]{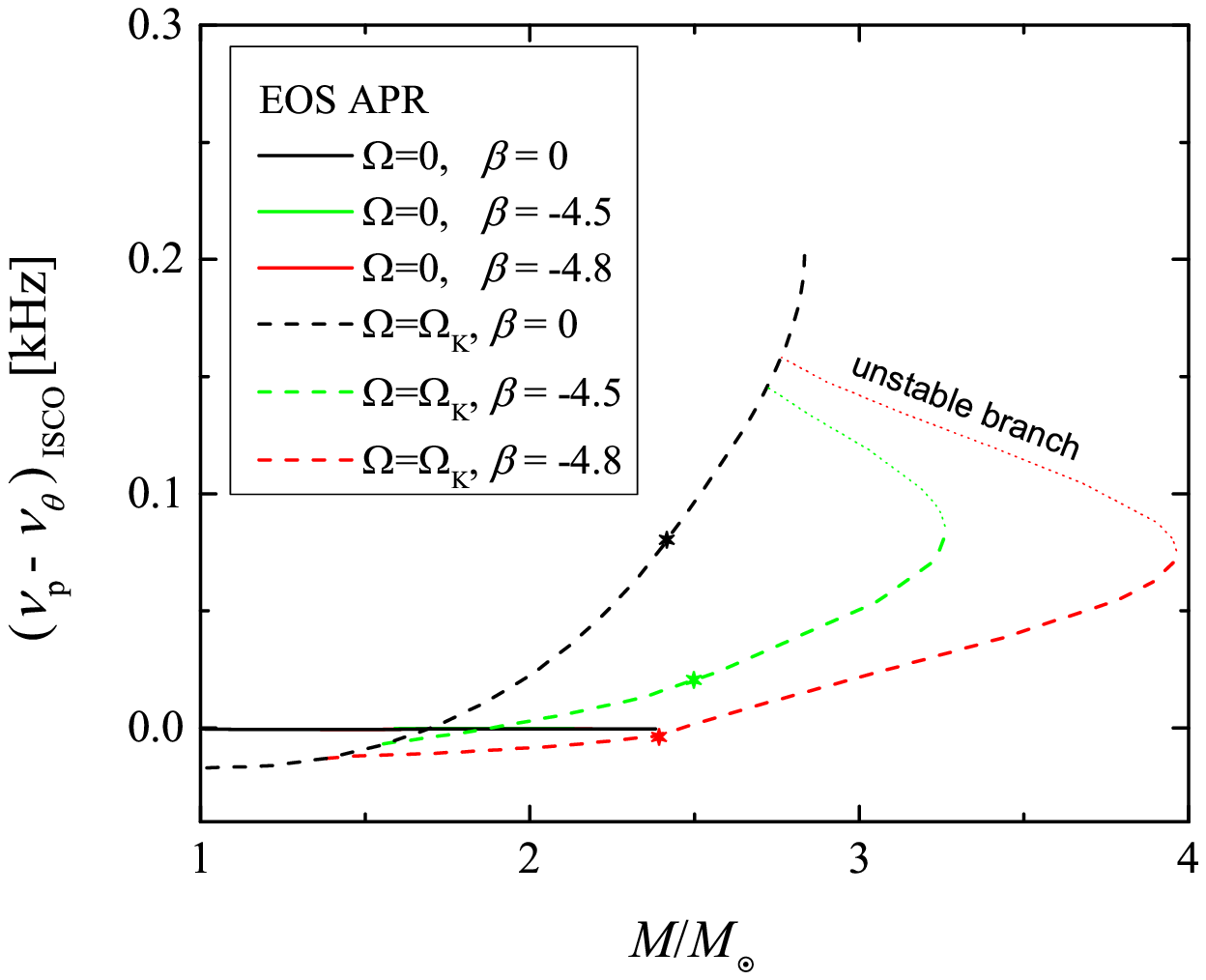}
\includegraphics[width=0.48\textwidth]{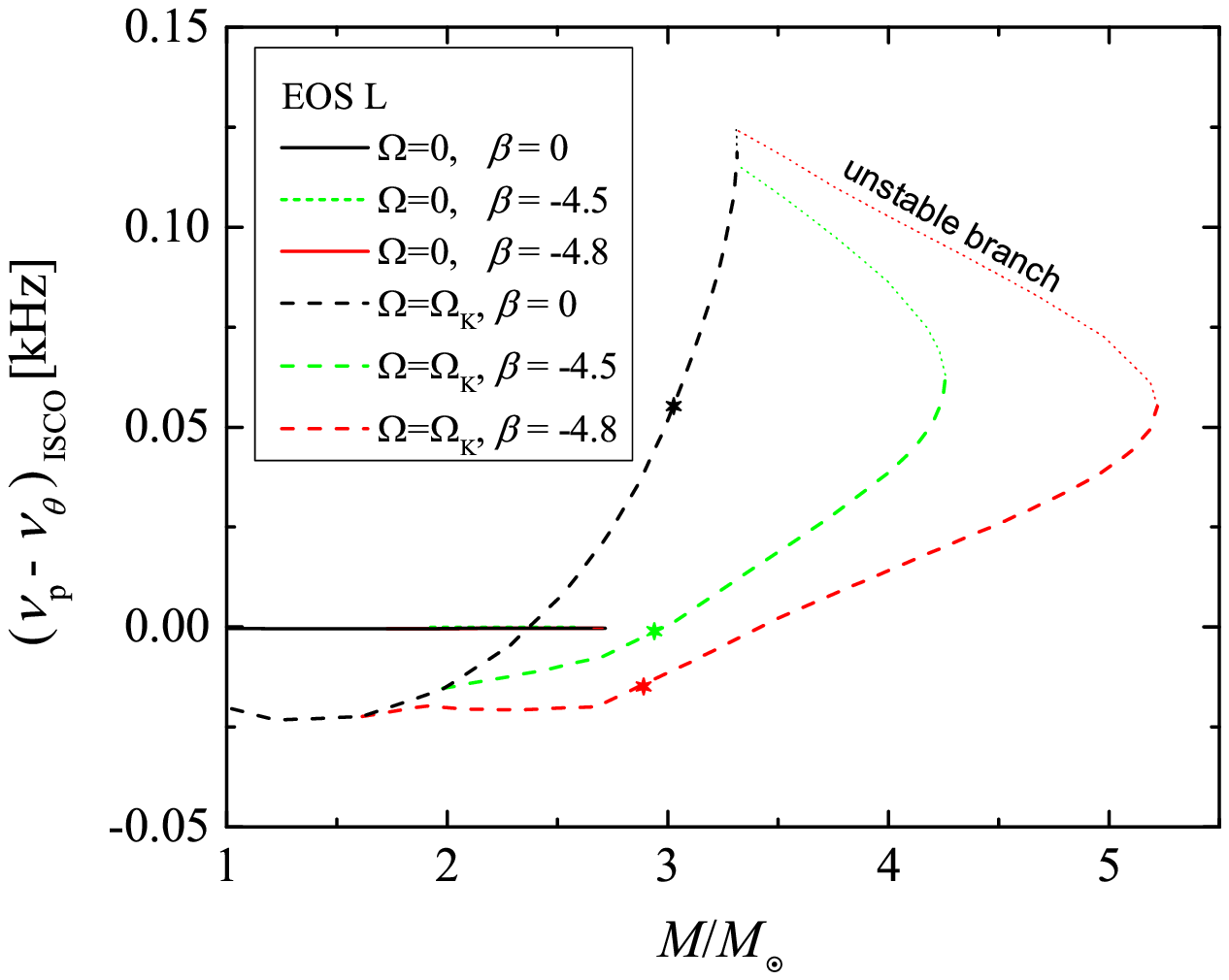}
\caption{The difference between the orbital frequency and the vertical epicyclic frequency at the ISCO as a function of mass. The notation is the same as in Fig. \ref{Fig:risco(M)}.}
\label{Fig:OmTh(M)}
\end{figure}

\begin{figure}[]
\centering
\includegraphics[width=0.48\textwidth]{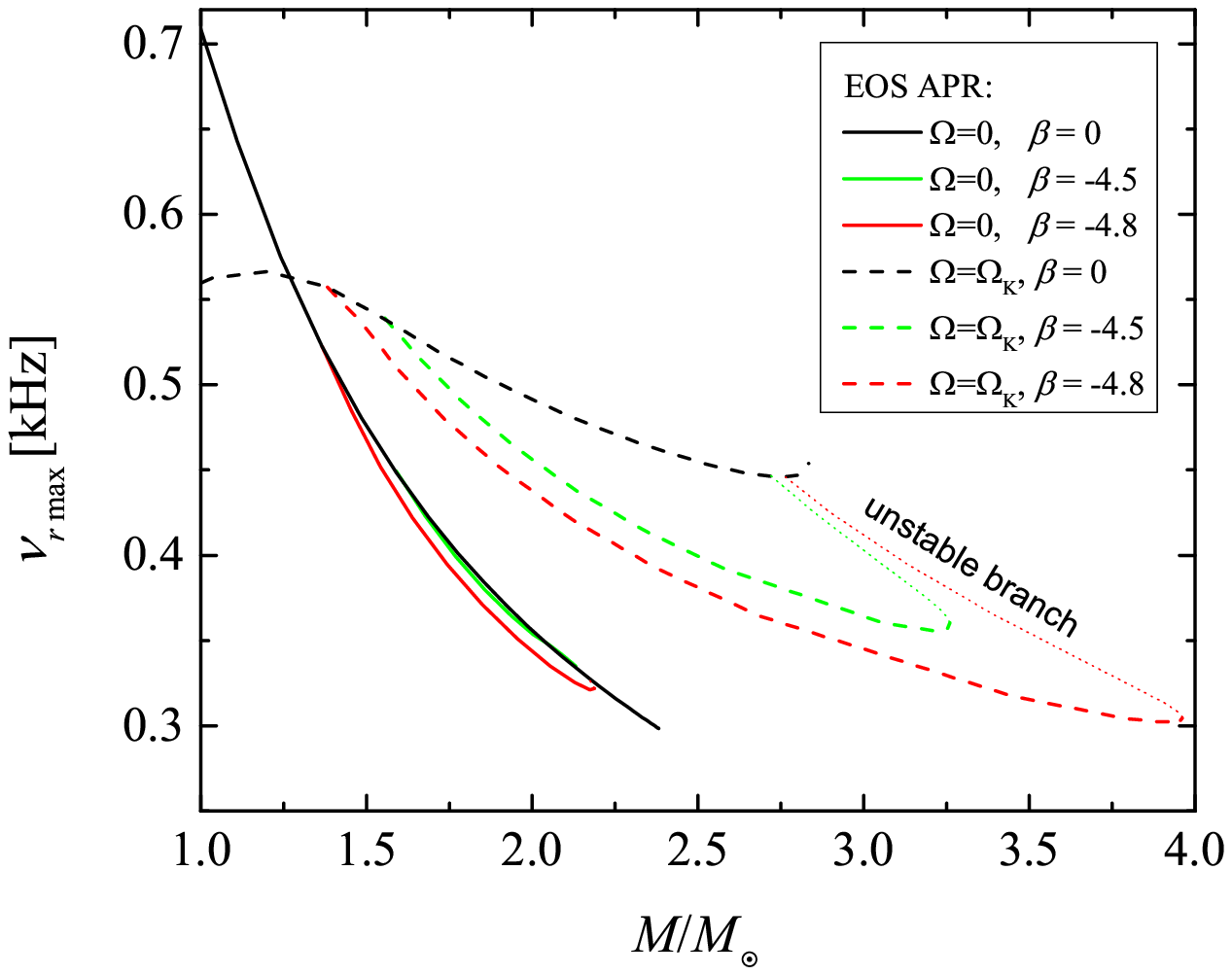}
\includegraphics[width=0.48\textwidth]{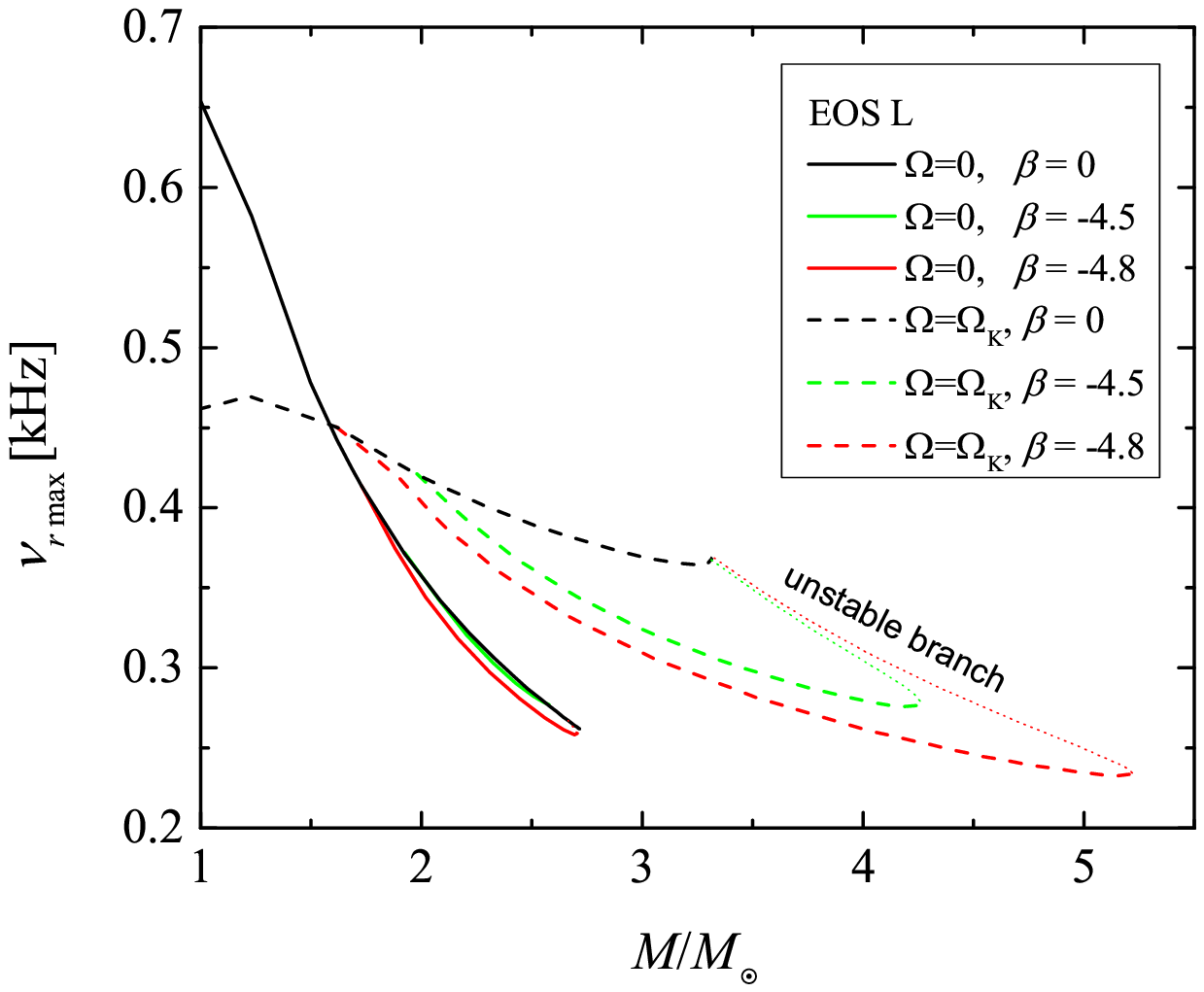}
\caption{Maximum value of the radial epicyclic frequency as a function of mass. The notation is the same as in Fig. \ref{Fig:risco(M)}.}
\label{Fig:omr1(M)}
\end{figure}

Figure \ref{Fig:OmK1(M)} shows the orbital frequency $\nu_p=\Omega_p/2\pi$ at the ISCO for nonrotating and masss-shedding sequences, as a function of mass. For the maximum mass models of the mass-shedding sequence, the orbital frequency at the ISCO decreases by 17\% for EOS\ APR and by 22\% for EOS L for STT models, compared to the corresponding models in GR.
Figure \ref{Fig:OmTh(M)} shows the difference between the orbital frequency $\nu_p$ and the vertical epicyclic frequency $\nu_\theta=\omega_\theta/2\pi$  at the ISCO. For nonrotating models this difference vanishes, while for models at the mass-shedding limit
this difference is reduced by 47\% for the maximum mass model of EOS APR and by 48\% for EOS L. Finally, Fig. \ref{Fig:omr1(M)} shows the corresponding plots for the maximum value  of the radial epicyclic frequency  $\nu_r=\omega_r/2\pi$, which is reduced by 22\% for the maximum mass model at the mass-shedding limit of EOS APR and by 26\% for EOS L.

In summary, while for nonrotating models the particular STT we consider has an almost negligible effect on both the structure and the orbital and epicyclic frequencies, the opposite is true for models at the mass-shedding limit, where differences to GR can become significant.

\begin{figure}[]
\centering
\includegraphics[width=0.48\textwidth]{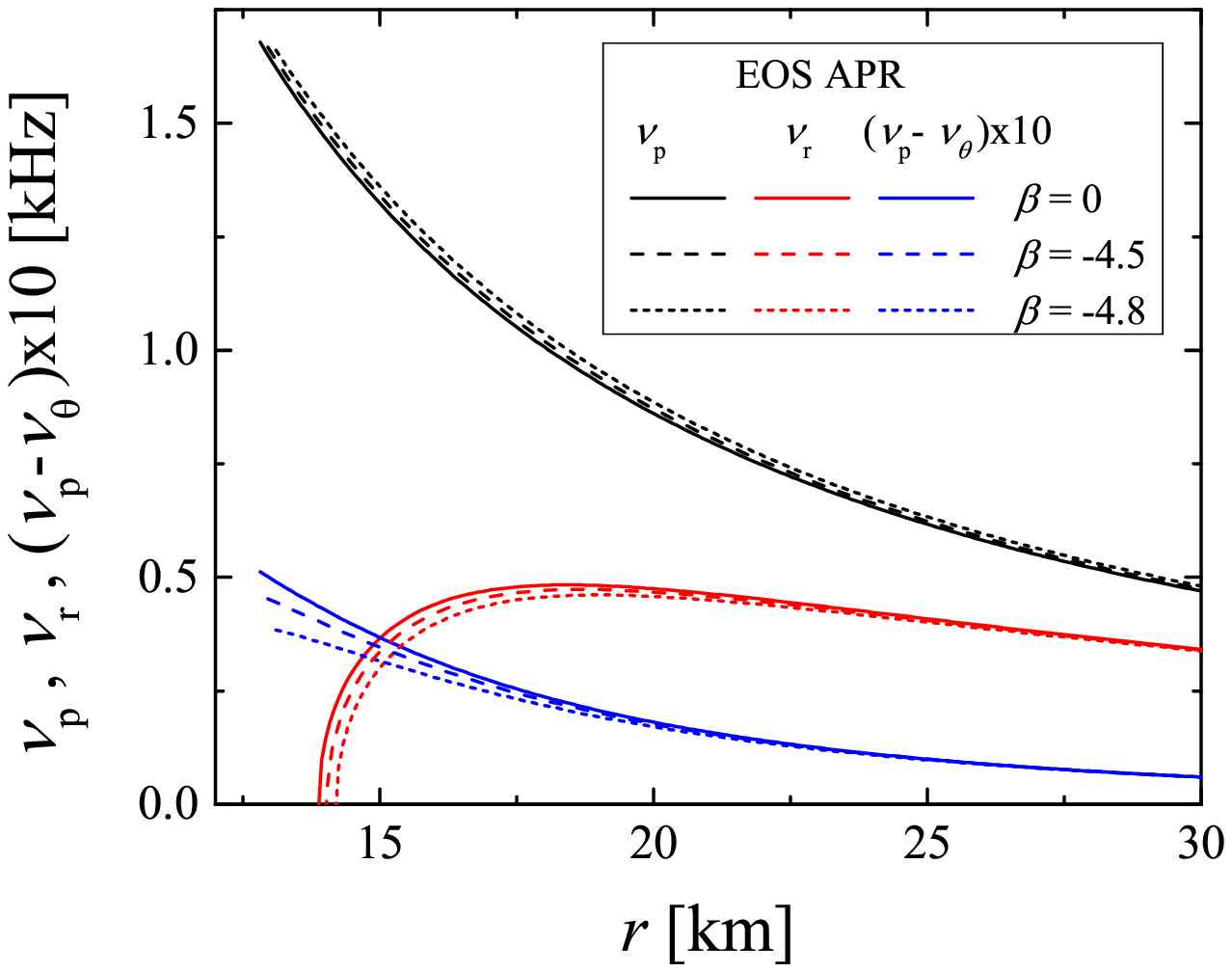}
\includegraphics[width=0.48\textwidth]{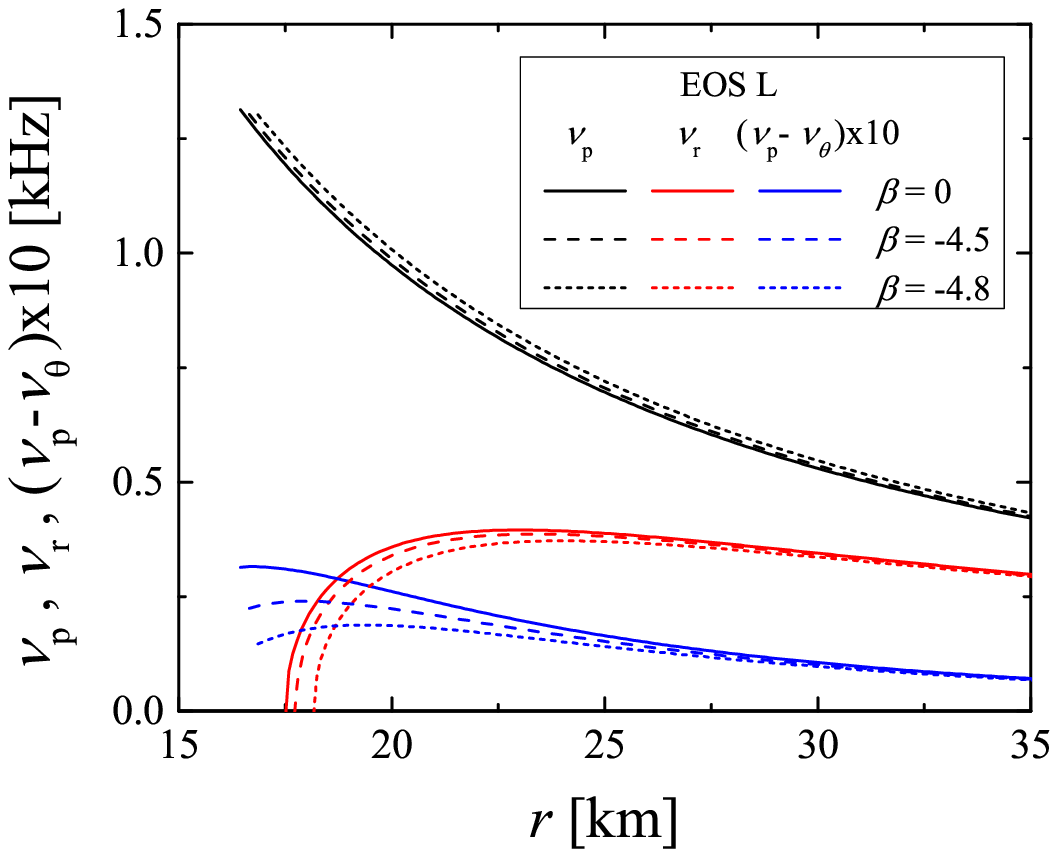}
\caption{Profiles of $\nu_p$, $\nu_r$ and $(\nu_p - \nu_\theta)\times 10$ outside the star, where $r$ is the circumferential radius of an orbit. The left panel considers a model with $M=1.8 M_\odot$, $\nu=700 Hz$ and EOS APR, and the right panel corresponds to $M=2.3 M_\odot$, $\nu=700 Hz$ and EOS L. Models with $\beta=-4.8$, $\beta=-4.5$ and $\beta=0$ are shown in each case. }
\label{Fig:omega(rc)_APR}
\end{figure}

For astrophysical observations, we consider two specific models of neutron stars rotating with (approximately) the maximum observed frequency for millisecond pulsars, $\nu=700Hz$ \cite{Hessels2006}. The two models are for EOS APR and EOS L, and they have masses $M=1.8 M_\odot$ and $M=2.3 M_\odot$ respectively\footnote{A larger mass is chosen for EOS L, because normally scalarization is stronger close to the maximum mass and the maximum mass for EOS L is quite high.}.  Fig. \ref{Fig:omega(rc)_APR} shows the profiles of the frequencies $\nu_p$, $\nu_r$ and $\nu_p - \nu_\theta $ as a function of radius .  The solid lines correspond to neutron stars in GR (with $\beta=0$) , while dashed lines are for the scalarized models with $\beta=-4.5$. For stable orbits (outside the ISCO) the differences between the GR and STT solutions are generally small. Only $\nu_p-\nu_\theta$ shows an appreciable difference at $R_{\rm ISCO}$ for the $M=2.3 M_\odot$\ model of EOS L.
In principle, this deviation could become important in models of quasi-periodic oscillations in low-mass x-ray binaries and could serve as a test of strong gravity (if other parameters are well constraint). Should more rapidly rotating neutron stars in LMXBs be discovered in the future, then stronger deviations are possible (up to the maximum deviations reported above for the maximum mass rotating models).

\subsection{Strange stars}
\begin{figure}[]
\centering
\includegraphics[width=0.48\textwidth]{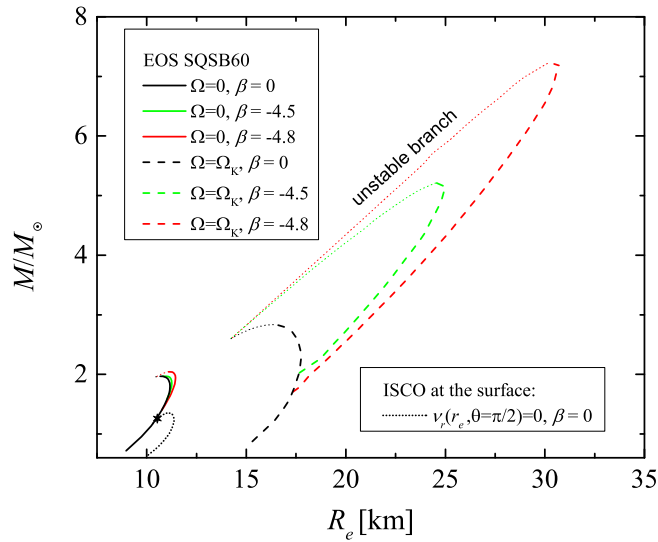}
\caption{Mass vs. radius relation for the strange star EOS SQSB60 for two different values of $\beta$. The solid lines correspond to nonrotating solutions and the dashed ones to models rotating at the mass-shedding limit.  The dotted line starting from the nonrotating GR solutions identifies the sequence of models where the corotating circular equatorial orbit at the stellar surface is marginally stable, i.e. models for which $\omega^{2}_r(r=r_e,\theta=\pi/2)=0$. The branches of equilibrium solutions above the maximum mass, which are unstable to collapse, are shown with a thin dotted line.}
\label{Fig:M(R_e)_SQS60}
\end{figure}

We consider the standard strange matter EOS with parameters described in \cite{Gondek-Rosinska2008} (denoted by SQSB60 there), for which the maximum mass for nonrotating model is very close to $2M_\odot$ (other versions of the strange star EOS can somewhat surpass this value, but quantitatively the results would be similar to our chosen EOS). Fig. \ref{Fig:M(R_e)_SQS60} shows the mass-radius relation for  nonrotating models and for models at the mass-shedding limit, for the same values of the coupling constant $\beta$ as for the hadronic EOSs in the previous section. To our knowledge, these are the first results for scalarized strange star models, even in the nonrotating case. The mass and radius of rapidly rotating models at the mass-shedding limit are dramatically larger than for the GR soution, much more than for hadronic EOSs. The maximum mass reaches $5M_\odot$, compared to $2.4 M_\odot$ in GR, with the corresponding radius reaching 25 km, compared to 17 km. Strange stars can reached a higher oblateness  compared to hadronic EOSs when rotating at the mass-shedding limit. Already, the GR solution deviates at mass-shedding more from the nonrotating models than the case for hadronic EOSs. Because the effect of scalarization on the equilibrium structure becomes more dramatic with increasing rotation \cite{Doneva2013}, the end result is a significantly larger deviation of STT solutions  from GR solutions than for hadronic EOSs.

Similar to the neutron star case, the ISCO for the less massive nonrotating strange star models in GR ($\beta=0$) coincides with the surface and only for the more massive models it is outside the stellar surface. But for all the models (above $1 M_\odot$ at least) rotating at the mass shedding limit, the ISCO is always located outside the star.
In Fig. \ref{Fig:M(R_e)_SQS60} the sequence of models for which the ISCO barely touches the surface is shown as a dotted line starting from the nonrotating sequence. The ISCO is above the stellar surface for all of the scalarized strange star models.
\textcolor[rgb]{0,0.537255,1}{}

\begin{figure}[]
\centering
\includegraphics[width=0.48\textwidth]{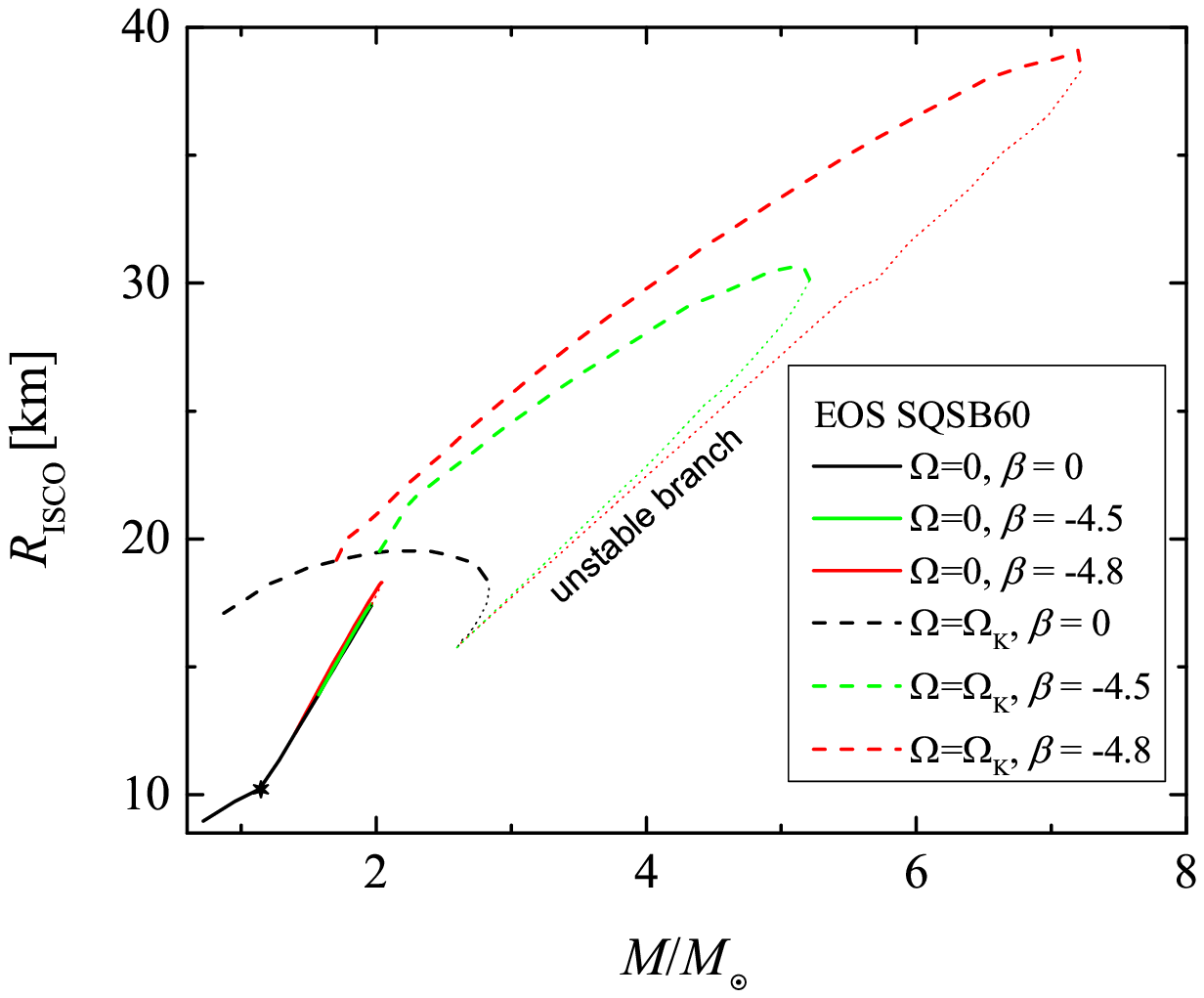}
\includegraphics[width=0.48\textwidth]{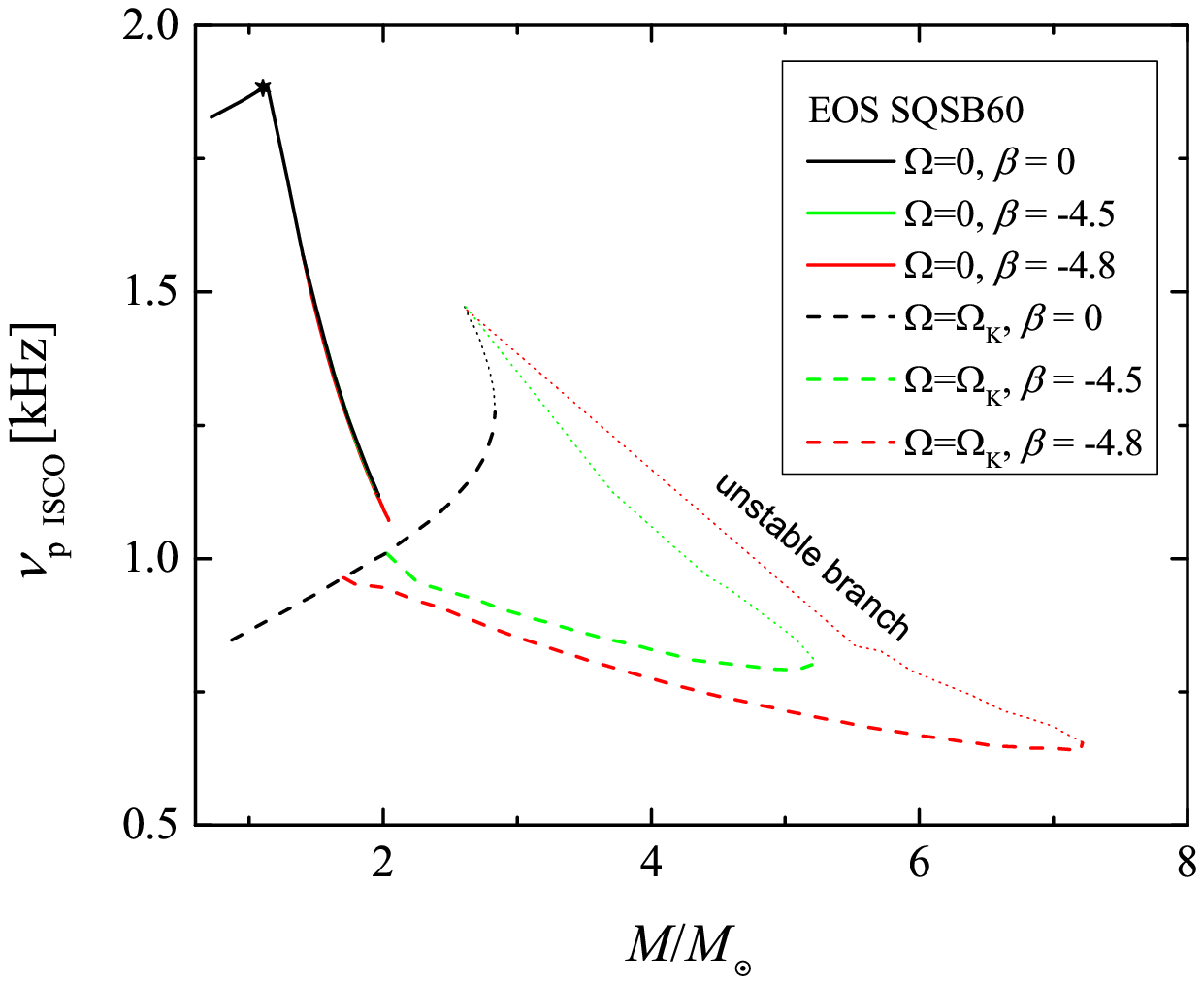}
\caption{{\it (Left panel)} The radius of the ISCO (or the stellar radius for low masses), as a function of mass. {\it (Right panel)} The Kepler frequency of a particle at the ISCO.}
\label{Fig:rISCO_OmK_SQS60}
\end{figure}

The left panel of Fig. \ref{Fig:rISCO_OmK_SQS60} shows the radius of the ISCO (or the radius of the star, when there are no unstable circular orbits), as a function of the mass for the nonrotating and mass-shedding sequences. For the maximum mass model on the mass-shedding sequence, the radius of the ISCO is  70\% larger than for the corresponding model in the GR case. The right panel of Fig. \ref{Fig:rISCO_OmK_SQS60} shows the orbital frequency at the ISCO along the nonrotating and mass-shedding sequences. In \cite{Stergioulas1999} it was found that for strange star models in GR, rotating at the mass-shedding limit, the orbital frequency at the ISCO deviates more from the nonrotating case than for hadronic EOSs, which can be seen in the right panel of Fig. \ref{Fig:rISCO_OmK_SQS60} for $\beta=0$. For the scalarized solutions, the orbital frequency decreases even more and for the maximum mass model at the mass-shedding limit it is 38\% smaller than in GR.

\begin{figure}[]
\centering
\includegraphics[width=0.48\textwidth]{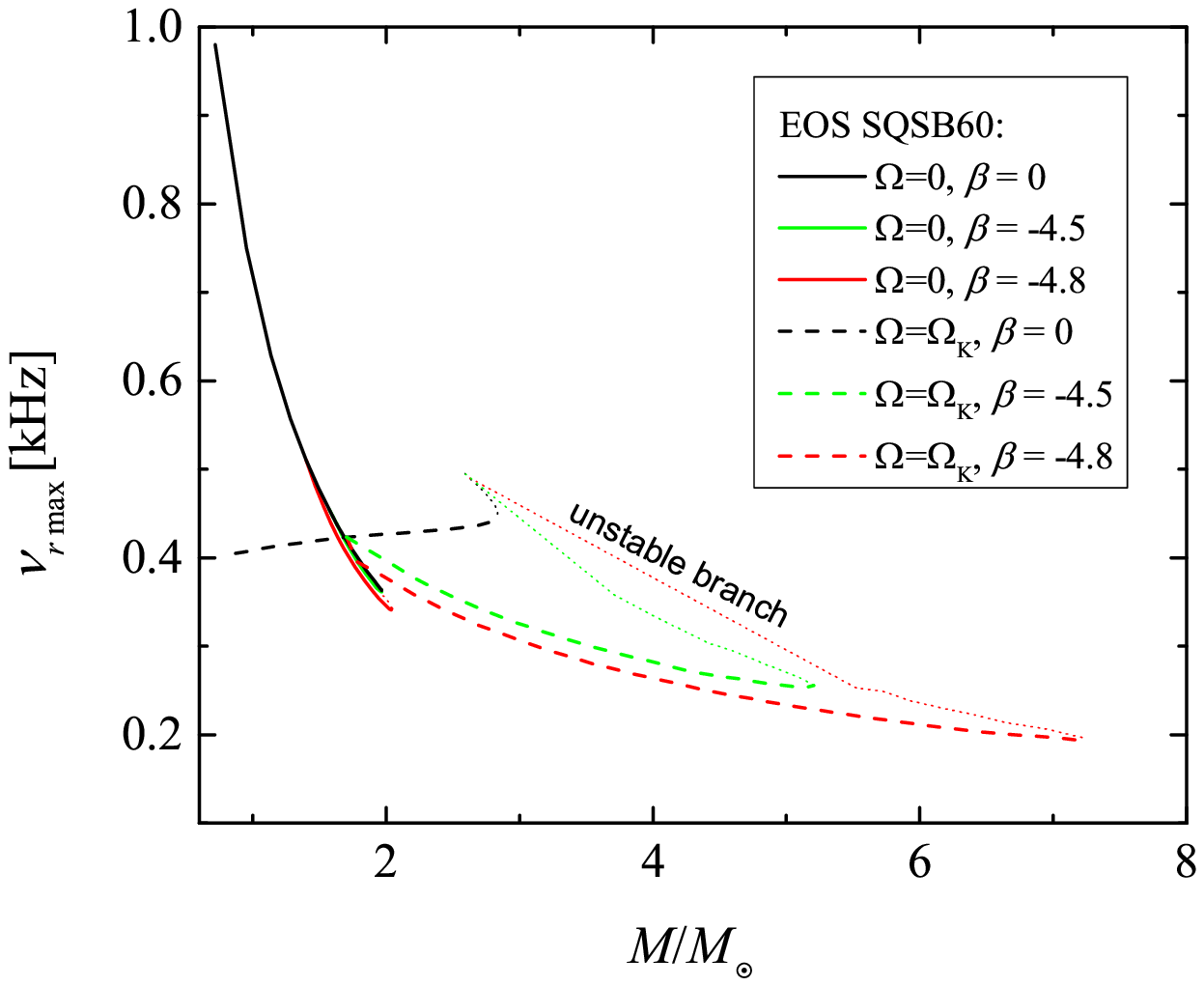}
\includegraphics[width=0.48\textwidth]{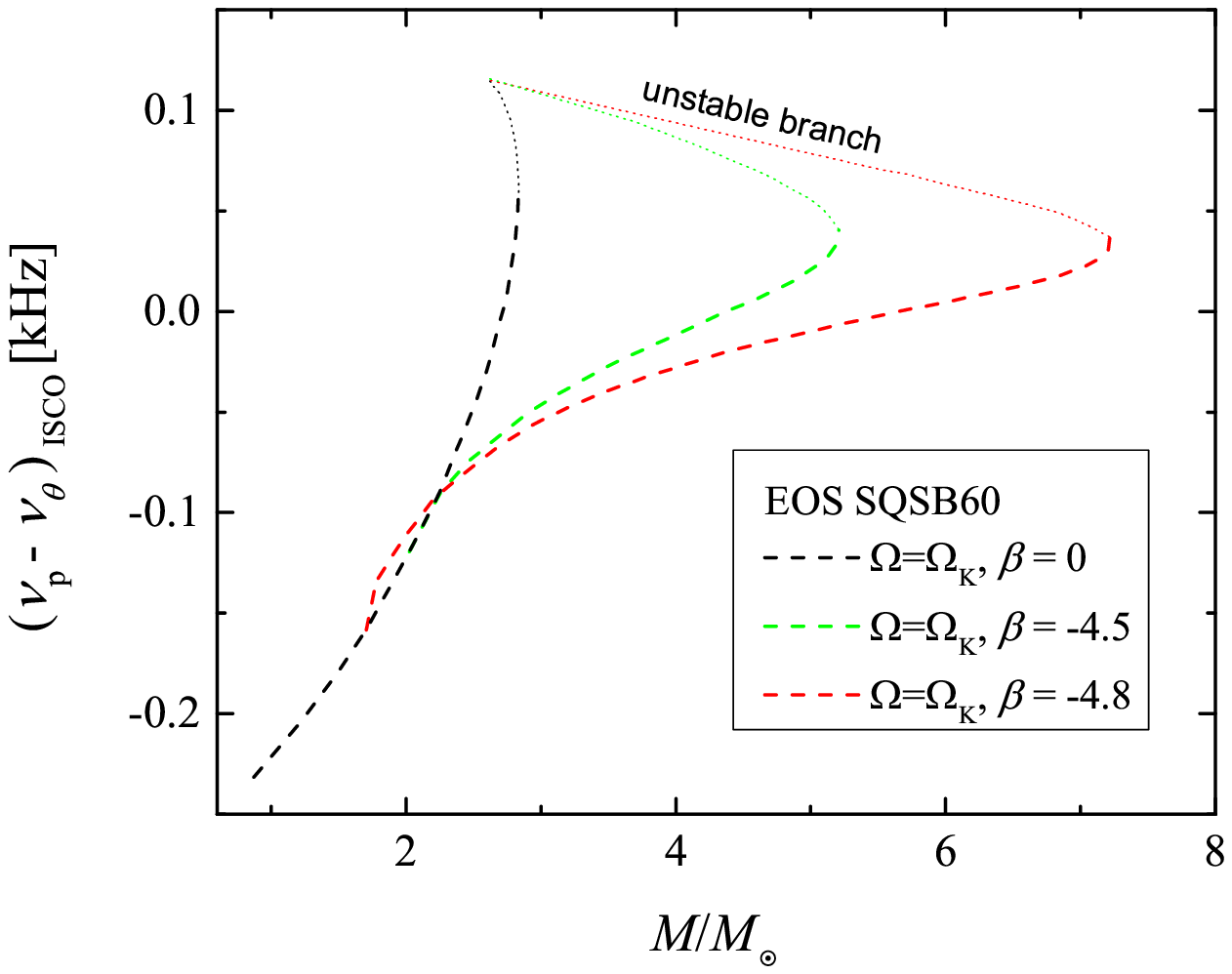}
\caption{{\it (Left panel)} The maximum value of the radial epicyclic frequency as a function of mass for strange stars. {\it (Right panel)} The difference between the orbital frequency and the vertical epicyclic frequency as a function of mass for strange stars rotating at the mass-shedding limit. }
\label{Fig:omr_omth_SQS60}
\end{figure}

The left panel of Fig. \ref{Fig:omr_omth_SQS60} shows the maximum value of the radial epicyclic frequency, which decreases by 38\% for the scalarized model with maximum mass at the mass-shedding limit, compared to the corresponding model in GR.
The right panel of Fig. \ref{Fig:omr_omth_SQS60}  shows that
$\nu_p - \nu_\theta$ for the maximum mass models on the mass-shedding limit has similar values in GR and STT, but for significantly different masses.

\begin{figure}[]
\centering
\includegraphics[width=0.48\textwidth]{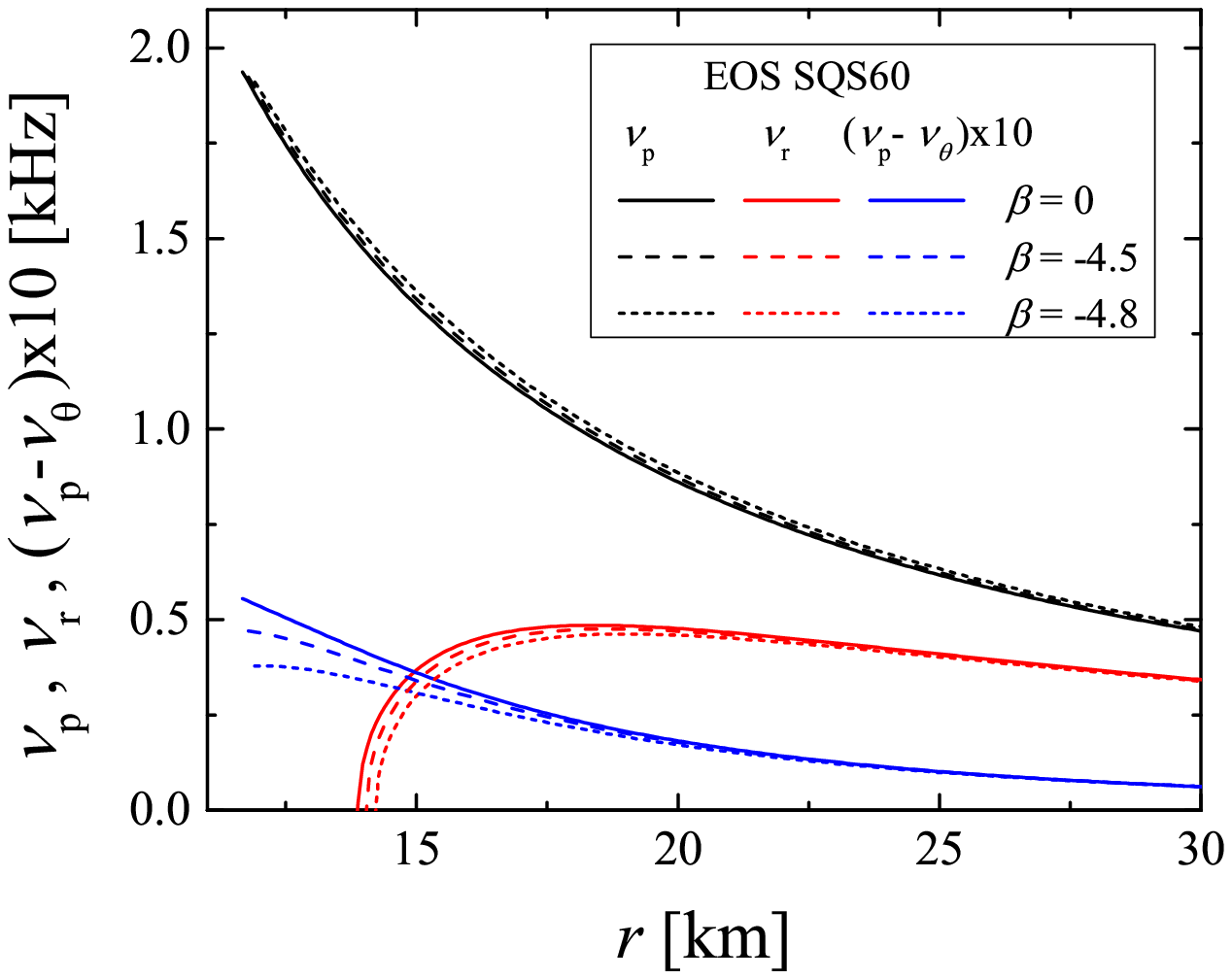}
\caption{The profiles of the quantities $\nu_p$, $\nu_r$ and $(\nu_p - \nu_\theta)\times 10$ outside the star, where $r$ is the  circumferential radius of an orbit, for scalarized strange stars ($\beta=-4.8$ and $\beta=-4.5$) and for a strange star with zero scalar field ($\beta=0$).
All models have the same mass of $M=1.8 M_\odot$ and rotational frequency of $\nu=700 Hz$.}
\label{Fig:omega(rc)_SQS60}
\end{figure}

 Finally, Fig. \ref{Fig:omega(rc)_SQS60} shows the profiles of $\nu_p$, $\nu_r$ and $\nu_p - \nu_\theta$ as a function of radius for $\beta=0, -4.5$ and $-4.8$. All three models are rotating with  a spin frequency of 700Hz and have a mass of $1.8M_\odot$. In this case, the rotation is too slow to significantly alter the above frequencies in STT, compared to GR.

\section{Conclusions}\label{Sec:Conclusions}
In the present paper we examined the orbital and epicyclic frequencies of particles orbiting rapidly rotating neutron stars within a particular class of scalar-tensor theories of gravity. Because geodesics formally only depend on the metric in scalar tensor theories, our formalism can be applied to any other class of STT. The interest in the orbital and epicyclic frequencies stems from the fact that these are involved in models that attempt to explain QPOs that are regularly observed in accretion disks.

We consider a class of STT that is indistinguishable from GR in the weak field regime, but it can differ significantly for strong fields. For this class of STT, scalarization of the solution in a  certain range of central densities is observed, i.e. the neutron stars can develop a strong, nontrivial scalar field. The effect of scalarization is more pronounced for rapidly rotating neutron stars.

The orbital and epicyclic frequencies of scalarized neutron stars were examined in  \cite{DeDeo2004}, but only for the case of nonrotating neutron stars. The extension to rapid rotation, considered here, is important for two reasons. First, neutron stars in LMXBs can reach high rotational frequencies, due to the accretion spin-up. Second, the developed nontrivial scalar field is much stronger for rapid rotation, compared to the static case. As a matter of fact, if we consider values of $\beta$ that are in agreement with the current observational constrains, only the rapidly rotating case could give us significant deviations from GR.

Using two representative hadronic EOSs and one strange star EOS, we studied in detail the effect of a  nontrivial scalar field on the orbital and epicyclic frequencies, along sequences of models that are nonrotating, rotating at the mass-shedding limit or rotating with a spin frequency of 700 Hz.   The effect of scalarization is marginal for nonrotating models, when one considers values of $\beta>-4.5$.   However, as the rotation increases, these differences become more appreciable, and at the mass-shedding limit, particles orbiting the scalarized compact stars have significantly different orbital and epicyclic frequencies than in GR. The position of the ISCO also changes considerably. The deviations from GR for strange star models can be larger compared to hadronic EOSs.

We studied in more detail neutron star models with masses around $2M_\odot$, and the maximum observed rotational frequency of $\sim 700{\rm Hz}$. It turns out that the presence of nontrivial scalar field is evident in the profiles of the orbital and epicyclic frequencies (even though EOS  uncertainties are comparable or  larger than this effect, at present). Should the EOS become tightly constraint in the future, then the current bound $\beta>-4.5$ allows of strong-field tests of alternative theories of gravity, especially in the case of a very stiff EOS, such as EOS L and for a sufficiently high mass.  If neutron stars in LMXBs spinning faster than 700 Hz are observed in the future, then it is  more likely that the orbital properties of particles in accretion disks could serve in strong-field gravity tests.

Notice that although our present paper focuses on orbital and epicyclic frequencies, we actually present here the first results for equilibrium configurations of rapidly rotating scalarized compact stars with either realistic hadronic EOS or strange matter EOS. The detailed investigation of the equilibrium properties of these object will be considered elsewhere, but from our current results we can make one important observation: the scalarized rapidly rotating strange stars can deviate stronger from the GR case in their equilibrium parameters, such as mass and radius, compared to neutron stars. The main reason is that the oblateness of strange stars can reach larger values at the mass-shedding limit, with respect to the neutron star case, which amplifies the effect of the scalar field.

For nonrotating models, we also provide tables of model parameters that allow the analytic construction of the exterior spacetime around a scalarized neutron star, as well as analytic expressions for  the orbital and radial epicyclic frequencies, which are useful for checking numerical codes.

\acknowledgments
DD would like to thank the Alexander von Humboldt Foundation for a stipend. KK and SY would like to thank the Research Group Linkage Programme of the Alexander
von Humboldt Foundation for the support. The support by the Bulgarian National Science Fund
under Grant DMU-03/6, by the Sofia University Research Fund under Grant 63/2014 and by the German Science Foundation (DFG) via
SFB/TR7 is gratefully acknowledged. Partial support comes from ``New-CompStar'', COST Action MP1304. NS is grateful for the hospitality of the T\"ubingen group during an extended visit.

\appendix

\section{ANALYTIC EXTERIOR\ METRIC FOR\ NONROTATING, SCALARIZED\ NEUTRON\ STARS}

 The exterior spacetime of a scalarized neutron star can be described analytically in the Einstein frame, using the Just metric [1959]
 \begin{equation}
ds_{*}^{2}=-\left(1- {a\over r} \right)^{b/a}dt^{2}+\left(1- {a\over r} \right)^{-b/a}[dr^{2}+(r^{2}-ar)(d\theta^{2}+sin^{2}\theta d\phi^{2})],
\end{equation}
where
\begin{equation}
b=2M
\end{equation}
($M$ is the mass, i.e. the ADM mass in the Einstein frame) and $a$ is obtained from
\begin{equation}
\omega_A^2=\frac{a^2-b^2}{4}.
\end{equation}
Here $\omega_A$ denotes the scalar charge.
The scalar field in the exterior is also given analytically, as
\begin{equation}
\phi (r)=\phi_{0}-\frac{\omega_A}{2M} \left(\frac{b}{a}\right) \ln \left(1-\frac{a}{r} \right)
\end{equation}
with $\phi_0$ being the asymptotic value of $\phi$ at infinity. In Tables \ref{Tab1} to \ref{Tab4} we provide the specifications for several nonrotating, scalarized models, constructed with four different equations of state. These tables can be used to obtain an analytic description for the exterior spacetime (both the metric and the scalar field).

Notice that the radial coordinate $r$ in the analytic Just metric does not coincide with the radial coordinate used in the quasi-isotropic form for rotating stars. In the GR limit, it coincides with the radial coordinate in Schwarzschild coordinates (while the quasi-isotropic form reduces to
isotropic coordinates). One can still compare directly quantities that do not depend on the definition of the radial coordinate (or one can do a full coordinate transformation).

{\renewcommand{\arraystretch}{1.2}
\begin{table} \label{Tbl:Coeff_APR}
\caption{Data for static models with EOS APR, $\beta=-4.5$. The displayed variables are as follows -- the central energy density in $[g/cm^3]$, the mass of the star in solar masses, the radius in  $[km]$, the value of the scalar field at the center of the star, the radius of ISCO in $[km]$, the orbital frequency of a particle at ISCO in $[kHz]$, the scalar charge $\omega_A$ measured in $[km]$, the value of the scalar field at the stellar surface and the values of the parameters $a$ and $b$ in the Just metric in $[km]$. }
\begin{centering}
\begin{tabular}{cccccccccc}
\hline
\hline
\noalign{\smallskip}
$\epsilon_c [g/cm^3]$ & $M/M_\odot$ & $R_e[km]$ & $\varphi_c$ & $R_{ISCO} [km]$ & $\nu_{p\;ISCO} [kHz]$ & $\omega_A [km]$ & $\varphi_s$ & $a [km]$ & $b [km]$  \\ \\
\hline
\noalign{\smallskip}
$9.50 \times 10^{14}$ & 1.68 & 12.22 &  -0.071 & 14.90 & 1.31 & -0.3596 & -0.0376 & 5.006 & 4.954 \\
$1.00  \times 10^{15}$ & 1.77 & 12.23 & -0.095 & 15.75 & 1.24 & -0.4868 & -0.0515 & 5.311 & 5.221 \\
$1.05  \times 10^{15}$ & 1.86 & 12.23 & -0.106 & 16.56 & 1.18 & -0.5448 & -0.0584 & 5.589 & 5.482 \\
$1.10  \times 10^{15}$ & 1.94 & 12.20 & -0.106 & 17.25 & 1.13 & -0.5415 & -0.0592 & 5.824 & 5.722 \\
$1.15  \times 10^{15}$ & 2.00 & 12.12 & -0.091 & 17.83 & 1.09 & -0.4608 & -0.0517 & 5.991 & 5.919 \\
$1.20  \times 10^{15}$ & 2.05 & 11.99 & -0.048 & 18.16 & 1.07 & -0.2414 & -0.0279 & 6.069 & 6.050 \\
\noalign{\smallskip}
\hline
\hline
\end{tabular}
\end{centering}
\label{Tab1}
\end{table}
}

{\renewcommand{\arraystretch}{1.2}
\begin{table} \label{Tbl:Coeff_L}
\caption{Data for static models with EOS L, $\beta=-4.5$. The column notations are the same as in Table I.}
\begin{centering}
\begin{tabular}{cccccccccc}
\hline
\hline
\noalign{\smallskip}
$\epsilon_c [g/cm^3]$ & $M/M_\odot$ & $R_e[km]$ & $\varphi_c$ & $R_{ISCO} [km]$ & $\nu_{p\;ISCO} [kHz]$ & $\omega_A [km]$ & $\varphi_s$ & $a [km]$ & $b [km]$  \\ \\
\hline
\noalign{\smallskip}
$5.95\times 10^{14}$ & 2.08 & 15.06 & -0.055 & 18.42 & 1.06 & -0.3467 & -0.0295 & 6.171 & 6.132 \\
$6.46\times 10^{14}$ & 2.20 & 15.09 & -0.094 & 19.62 & 1.00 & -0.6004 & -0.0516 & 6.620 & 6.510 \\
$6.97\times 10^{14}$ & 2.32 & 15.10 & -0.111 & 20.72 & 0.94 & -0.7074 & -0.0616 & 7.010 & 6.866\\
$7.49\times 10^{14}$ & 2.43 & 15.07 & -0.113 & 21.63 & 0.90 & -0.7176 & -0.0638 & 7.313 & 7.171 \\
$8.00\times 10^{14}$ & 2.51 & 14.96 & -0.101 & 22.28 & 0.88 & -0.6299 & -0.0574 & 7.507 & 7.400 \\
$8.49\times 10^{14}$ & 2.55 & 14.78 & -0.068 & 22.64 & 0.86 & -0.4162 & -0.0391 & 7.580 & 7.535 \\
\noalign{\smallskip}
\hline
\hline
\end{tabular}
\end{centering}
\label{Tab2}
\end{table}
}

{\renewcommand{\arraystretch}{1.2}
\begin{table} \label{Tbl:Coeff_Sly4}
\caption{Data for static models with EOS SLy4, $\beta=-4.5$. The column notations are the same as in Table I.}
\begin{centering}
\begin{tabular}{ccccccccccc}
\hline
\hline
\noalign{\smallskip}
$\epsilon_c [g/cm^3]$ & $M/M_\odot$ & $R_e[km]$ & $\varphi_c$ & $R_{ISCO} [km]$ & $\nu_{p\;ISCO} [kHz]$ & $\omega_A [km]$ & $\varphi_s$ & $a [km]$ & $b [km]$  \\ \\
\hline
\noalign{\smallskip}
$1.10\times 10^{15}$ & 1.53 & 11.63 & -0.045 & 13.57 & 1.44 & -0.2083 & -0.0226 & 4.542 & 4.523 \\
$1.20\times 10^{15}$ & 1.62 & 11.58 & -0.090 & 14.46 & 1.35 & -0.4208 & -0.0466 & 4.872 & 4.798 \\
$1.30\times 10^{15}$ & 1.72 & 11.52 & -0.107 & 15.30 & 1.28 & -0.5000 & -0.0566 & 5.165 & 5.067 \\
$1.40\times 10^{15}$ & 1.80 & 11.43 & -0.106 & 15.98 & 1.22 & -0.4917 & -0.0572 & 5.394 & 5.303 \\
$1.50\times 10^{15}$ & 1.85 & 11.28 & -0.083 & 16.47 & 1.18 & -0.3772 & -0.0454 & 5.526 & 5.474 \\
$1.55\times 10^{15}$ & 1.87 & 11.18 & -0.055 & 16.58 & 1.18 & -0.2465 & -0.0303 & 5.547 & 5.525 \\
\noalign{\smallskip}
\hline
\hline
\end{tabular}
\end{centering}
\label{Tab3}
\end{table}
}

{\renewcommand{\arraystretch}{1.2}
\begin{table} \label{Tbl:Coeff_SQS60}
\caption{Data for static models with EOS SQS60, $\beta=-4.5$. The column notations are the same as in Table I.}
\begin{centering}
\begin{tabular}{cccccccccc}
\hline
\hline
\noalign{\smallskip}
$\epsilon_c [g/cm^3]$ & $M/M_\odot$ & $R_e[km]$ & $\varphi_c$ & $R_{ISCO} [km]$ & $\nu_{p\;ISCO} [kHz]$ & $\omega_A [km]$ & $\varphi_s$ & $a [km]$ & $b [km]$  \\ \\
\hline
\noalign{\smallskip}
$9.00\times 10^{14}$ & 1.63 & 11.08 & -0.029 & 14.48 & 1.35 & -0.1488 & -0.0176 & 4.833 & 4.824 \\
$1.00\times 10^{15}$ & 1.72 & 11.18 & -0.081 & 15.31 & 1.28 & -0.4137 & -0.0489 & 5.151 & 5.084 \\
$1.10\times 10^{15}$ & 1.79 & 11.24 & -0.101 & 15.99 & 1.22 & -0.5145 & -0.0611 & 5.396 & 5.297 \\
$1.20\times 10^{15}$ & 1.85 & 11.26 & -0.110 & 16.50 & 1.19 & -0.5578 & -0.0668 & 5.579 & 5.466 \\
$1.30\times 10^{15}$ & 1.89 & 11.25 & -0.113 & 16.91 & 1.16 & -0.5644 & -0.0683 & 5.709 & 5.596 \\
$1.40\times 10^{15}$ & 1.93 & 11.21 & -0.111 & 17.16 & 1.14 & -0.5434 & -0.0666 & 5.795 & 5.692 \\
$1.50\times 10^{15}$ & 1.95 & 11.15 & -0.104 & 17.35 & 1.13 & -0.4987 & -0.0620 & 5.843 & 5.758 \\
$1.60\times 10^{15}$ & 1.96 & 11.08 & -0.092 & 17.44 & 1.12 & -0.4306 & -0.0545 & 5.860 & 5.796 \\
$1.70\times 10^{15}$ & 1.97 & 10.98 & -0.073 & 17.46 & 1.12 & -0.3332 & -0.0429 & 5.849 & 5.811 \\
$1.80\times 10^{15}$ & 1.97 & 10.88 & -0.039 & 17.43 & 1.12 & -0.1738 & -0.0228 & 5.816 & 5.806 \\
\noalign{\smallskip}
\hline
\hline
\end{tabular}
\end{centering}
\label{Tab4}
\end{table}
}

\section{ANALYTIC\ EXPRESSIONS\ FOR\ THE\ ORBITAL\ AND\ EPICYCLIC\ FREQUENCIES}

\

Using the effective potential approach (see Abramowicz [2004])\ we find the following analytic expression for the orbital frequency:
\begin{equation}
\Omega_{p}^{2}=\frac{(1-a/r)^{2b/a}[ab+2\beta d^{2}\ln(1-a/r)]}{(a-r)r[a(a+b-2r)-2\beta d^{2}\ln(1-a/r)]},
\end{equation}
which has the expected GR limit of
\begin{equation}
\Omega_{p}^{2}=M/r^{3}.
\end{equation}
We also obtain the following analytic expression for the radial epicyclic frequency:
\begin{eqnarray}
 \omega_{r}^2 &=& -\-\left(1-\frac{a}{r}\right)^{2 b/a}  \left\{2 a r^2 (a-r)^2 \left[ a (a+b-2 r)-2 \beta  d^2 \ln \left(1-\frac{a}{r}\right)\right]    \right \}^{-1} \nonumber \\
   &&\times  \left \{ a^2 \left[
   (a+2 b) \left(b (a+b)-2 \beta
    d^2\right)
       -2 r \left(b (a+3 b)-2 \beta  d^2\right)
    +2 b r^2 \right] \right. \nonumber\\
    &&+ \left.2 \beta  d^2 \ln \left(1-\frac{a}{r}\right) \left[ a \left(a^2+2 a (b-r)+2 r (r-2 b)\right)-2 \beta  d^2
   (a+2 b-2 r) \ln \left(1-\frac{a}{r}\right)\right]\right\},
\end{eqnarray}
which has the expected GR\ limit of
\begin{equation}
\omega_{r}^2=\frac{M(r-6M)}{r^{4}}.
\end{equation}
Using the above analytic expressions, we have confirmed that our numerical code has a high accuracy  in obtaining the radial epicyclic frequencies of scalarized, nonrotating stars (we compared the orbital frequency at the location of the ISCO, i.e. where $\omega_{r}$ goes to zero).


\bibliography{references}

\end{document}